\documentclass{aa}  

\usepackage{graphicx}
\usepackage{txfonts}
\usepackage{placeins}
\usepackage{natbib}
\usepackage{caption}
\usepackage{subfigure}
\usepackage{color}
\usepackage{stfloats}
\usepackage{float}
\usepackage{threeparttable}
\usepackage{amsmath}
\usepackage{adjustbox}
\usepackage[colorlinks=true,     linkcolor=blue, citecolor=blue, filecolor=blue, urlcolor=blue]{hyperref}
\begin{document}

\title{Magnetic field of the roAp star KIC~10685175: observations versus theory}
\subtitle{}

\author{Fangfei Shi
          \inst{1,2}
          \and
          Huawei Zhang\inst{1,2}
          \and
          Swetlana Hubrig\inst{3}
          \and
          Silva J{\"a}rvinen\inst{3}
          \and
          Huiling Chen\inst{1,2}
          \and
          Tianqi Cang\inst{4}
          \and
          Jianning Fu\inst{4,5}
          \and
          Donald Kurtz\inst{6,7}
          }
          
 \institute{Department of Astronomy, Peking University, Beijing 100871, China\\
              \email{fangfei1420@pku.edu.cn,zhanghw@pku.edu.cn}
         \and
             The Kavli Institute for Astronomy and Astrophysics, Peking University, Beijing 100871, China\\
            \and
             Leibniz-Institut f{\"u}r Astrophysik Potsdam (AIP), An der Sternwarte 16, 14482 Potsdam, Germany\\
             \and
             School of Physics and Astronomy, Beijing Normal University, Beijing 100875, China\\
             \and
             Institute for Frontiers in Astronomy and Astrophysics, Beijing Normal University, Beijing 102206, China\\
              \and
             Centre for Space Research, North-West University, Mahikeng 2745, South Africa\\
              \and
             Jeremiah Horrocks Institute, University of Central Lancashire, Preston PR1 2HE, UK\\         
             }

\abstract
{KIC~10685175 is a roAp star whose polar magnetic field is predicted to be 6\,kG through a non-adiabatic axisymmetric pulsation theoretical model. }
{In this work, we aim to measure the magnetic field strength of KIC~10685175 using high-resolution spectropolarimetric observations, and compare it with the one predicted by the theoretical model.}
  %We present the magnetic field measurements for KIC~10685175 to compare the observed and the predicted magnetic field.}
{Two high-resolution unpolarized spectra have been analyzed to study for the presence of the magnetically split lines and derive the [Fe/H] ratio of
  this star through equivalent width measurements of ten Fe lines. One polarized spectrum has been used to measure the mean longitudinal
  magnetic field with the Least-Squares Deconvolution technique. Further, to examine the presence of chemical spots on the stellar surface, we
  have measured the mean longitudinal magnetic fields using different lines belonging to different elements.}
{From the study of two high-resolution unpolarized spectra, we obtained [$T_{\rm eff}$, $\log g$, [Fe/H], [$\alpha$/Fe], $V_{mic}$]=[8250 $\pm$ 200\,K, 4.4 $\pm$ 0.1, -0.4 $\pm$ 0.2, 0.16 $\pm$ 0.1, 1.73 $\pm$ 0.2\,km~s$^{-1}$].  Although the Fe absorption lines appear relatively weak in
  comparison to typical  Ap stars with similar $T_{\rm eff}$, the lines belonging to rare earth elements (Eu and Nd) are stronger than that in chemically normal stars,
  indicating the peculiar nature of KIC~10685175.
  The mean longitudinal magnetic field $\langle B_\ell \rangle=-226\pm39$\,G has been measured in the polarized spectrum, but magnetically split lines
  were not detected. No significant line profile variability was evident in our spectra. Also the
    longitudinal magnetic field strengths measured using line masks constructed for different elements have been rather similar. Due to a poor
    rotation phase coverage of our data, additional spectroscopic and polarimetric observations are needed to
allow us to conclude on the inhomogeneous element distribution over the stellar surface.}
{The estimated polar magnetic field is $4.8 \pm 0.8$\,kG, which is consistent
  with the predicted polar magnetic field strength of about 6\,kG within 3$\sigma$. This work therefore provides support for the pulsation
  theoretical model.}


\maketitle
\nolinenumbers{
\section{Introduction}           
%\label{sect:intro}

Ap stars are chemically peculiar (CP) stars of spectral types A which have overabundances of iron-peak and rare-earth elements, such as Si, Cr, Sr, and Eu \citep{1974ARA&A..12..257P}.}
A prime factor governing the peculiarities of Ap stars is the strong magnetic fields they host. These
 magnetic fields are roughly dipolar with typical field strengths of a kG, or more. The strong magnetic field suppresses convection, thus
 providing a stable environment for ions with many absorption features to be lifted by radiation pressure against gravity and kept stratified in the observable
 surface layers. At the same time, some other elements, especially helium, that have few absorption lines near flux maximum in
 the outer envelope of the star settle down affected by gravity. The element stratification in the presence of a magnetic field causes the peculiarities of surface abundance.

 Some Ap stars exhibit high-overtone, low-degree pressure pulsation modes. These are called rapidly oscillating Ap (roAp) stars and they
 have many pulsation features. In the HR diagram, these stars overlap with $\delta$ Sct stars which are non-magnetic (or extremely weakly
 magnetic). Generally, the $\delta$ Sct stars pulsate in low overtone radial p~modes while the roAp stars pulsate in high radial overtone
 magneto-acoustic modes. Also, unlike the non-magnetic stars, the pulsation axes of roAp stars are almost symmetric about the oblique quasi-dipolar
 magnetic field \citep{2003A&A...404..669K,2005MNRAS.360.1022S,2011A&A...536A..73B}. Thus it is believed that the strong magnetic fields of roAp
 stars play an essential role in the pulsation mode excitation and selection \citep{2001MNRAS.323..362B,2005MNRAS.360.1022S,2006MNRAS.365..153C}.

The pulsation axes of roAp stars are affected by the presence of a magnetic field as discussed in \citet{1996ApJ...458..338D}, \citet{2000MNRAS.319.1020C}, and
 \citet{2004MNRAS.350..485S}, which means the pulsation mode is not a pure dipole or quadrupole (or higher).
 \cite{2005MNRAS.360.1022S} provides a method to model the quadrupole or dipole pulsation distorted by dipole magnetic fields. In this model,
 the polar magnetic field is predicted by comparing the theoretical pulsation data with the observed one.
 Further, this model is used to estimate the polar magnetic field strength, effective temperature, mass, and other stellar parameters \citep{2016MNRAS.462..876H,2018MNRAS.476..601H}.
 %For example, \citet{2018MNRAS.480.2405H} derived the effective temperature and the luminosity of TYC 2488-1241-1 and found that its observed frequencies are above the theoretical acoustic cutoff frequency. 
% The polar magnetic fields of many roAp stars have been estimated, 
 %For example, for stars 2MASS~16400299-0737293 (1.7\,kG, \citealt{2018MNRAS.476..601H}) and HD~24355 (1.4\,kG, \citealt{2016MNRAS.462..876H}). Moreover, the authors also estimated the mean magnetic field strength of 2.64\,kG for through high-resolution spectra, but did not compare the predicted and the measured magnetic field.
On the other hand, a test of consistency between the field measurements and the predictions on the magnetic field strengths permits to support the theoretical model, or place new constraints on it.

 KIC~10685175 is a roAp star that has been observed by both {\it Kepler} \citep{2010Sci...327..977B,2010ApJ...713L..79K} and and the Transiting Exoplanet Survey Satellite \citep[TESS;][]{2015JATIS...1a4003R}. \cite{2016AJ....151...13G} first classified KIC~10685175 as an Ap star with spectral type A6 IV (Sr)Eu using a LAMOST \citep{2012RAA....12..723Z,2012RAA....12.1197C} low-resolution spectrum, and a second classification of A4V\,Eu was given by \cite{H_mmerich_2018} using a high-resolution spectrum obtained at Star\'a Lesn\'a Observatory. However,
  magnetic field measurements have not been carried out for this star yet. 
  \cite{2020ApJ...901...15S} studied the rotation and pulsation features of this star. Using their values for the rotation period, $P_{\rm rot} = 3.1028$\,d, and the estimated radius, $R = 1.39\pm0.05$\,R$_{\odot}$ from \citet{2018AA...616A...1G}, we obtain the equatorial rotation velocity $v_{\rm eq} = 22.7\pm0.8$~km~s$^{-1}$. Assuming the inclination $i= 60^{\circ}$ from the work of \cite{2020ApJ...901...15S} we get $v\sin i = 19.6\pm0.7$~km~s$^{-1}$. The uncertainty of $\sin i$ is not considered since it cannot be derived through the pulsation theoretical model in \cite{2020ApJ...901...15S}. From the theoretical model presented by \cite{2005MNRAS.360.1022S}, the predicted dipole magnetic field strength is about 6\,kG, which is expected to be detectable in a high-resolution spectra. If this predicted magnetic field can be confirmed by observations, this star will belong to a group of roAp stars with rather strong surface magnetic fields.

There are two main observational strategies to measure the magnetic field: one is related to spectroscopic observations of the magnetically split lines and observations of circular polarization. High-resolution spectroscopic observations permit to determine the average surface magnetic field, $\langle B \rangle$, by measuring the splitting between the Zeeman components in magnetically sensitive lines.
 This method is simple and straightforward, but requires high resolution and signal-to-noise ratio (SNR).
 Another method to measure magnetic fields is high-resolution spectropolarimetric observations.
 % Magnetic fields also cause polarization signatures in spectral lines through the Zeeman effect.
 This method allows us to measure the average over the stellar
 disc of the line-of-sight component of the magnetic vector, which is called the mean longitudinal magnetic field.
The observations of circular polarization are the most widely used to measure magnetic fields in Ap star and it does not need very high SNR.
%In addition, linear polarization can also be used to measure the transverse magnetic field \citep{1995A&AS..114...79L}, but since linear polarization is more difficult to obtain, few works apply this method.

\section{Observations and data reduction}

The spectra were obtained using the Echelle Spectropolarimetric Device for the Observation of Stars (ESPaDOnS) on the 3.6-m
Canada-France-Hawaii Telescope (CFHT). ESPaDOnS is a bench-mounted high-resolution echelle spectrograph and spectropolarimeter with
wavelength coverage from 370 to 1050 nm \citep{2006ASPC..358..362D}. There are two observation modes for ESPaDOnS -- spectropolarimetric mode with a resolving power of $R \sim 68\,000$ and spectroscopic mode with higher resolution up to $R \sim 81\,000$.

Two high-resolution ($R\sim81\,000$) spectra were obtained in spectroscopic mode on 2021 September 1. Using the exposure time of 1300\,s
for each spectrum we achieved the SNR of about 60 at 6120\,$\rm \AA$.  On 2022 October 20, a lower resolution ($R\sim68\,000$) circularly polarised
spectrum in spectropolarimetric mode was obtained as a follow-up observation. The spectropolarimetric observation consisted of four
sub-exposures observed at different positions of the quarter-wave retarder plate. Each sub-exposure time is 1600~s. The achieved SNR of
integral light spectra at 6120\,$\rm \AA$ is 90.
The data was reduced using Libre-ESpRIT \citep{1997MNRAS.291..658D}, which is the pipeline built for the ESPaDOnS. The reduction
includes optimal spectrum extraction and normalization.
More detailed information on the observations is presented in Table~\ref{tab:obs}. 

\begin{figure}[htb]
%\center{\includegraphics[width=13cm]  {compare.png}} 
\center{\includegraphics[width=8cm]  {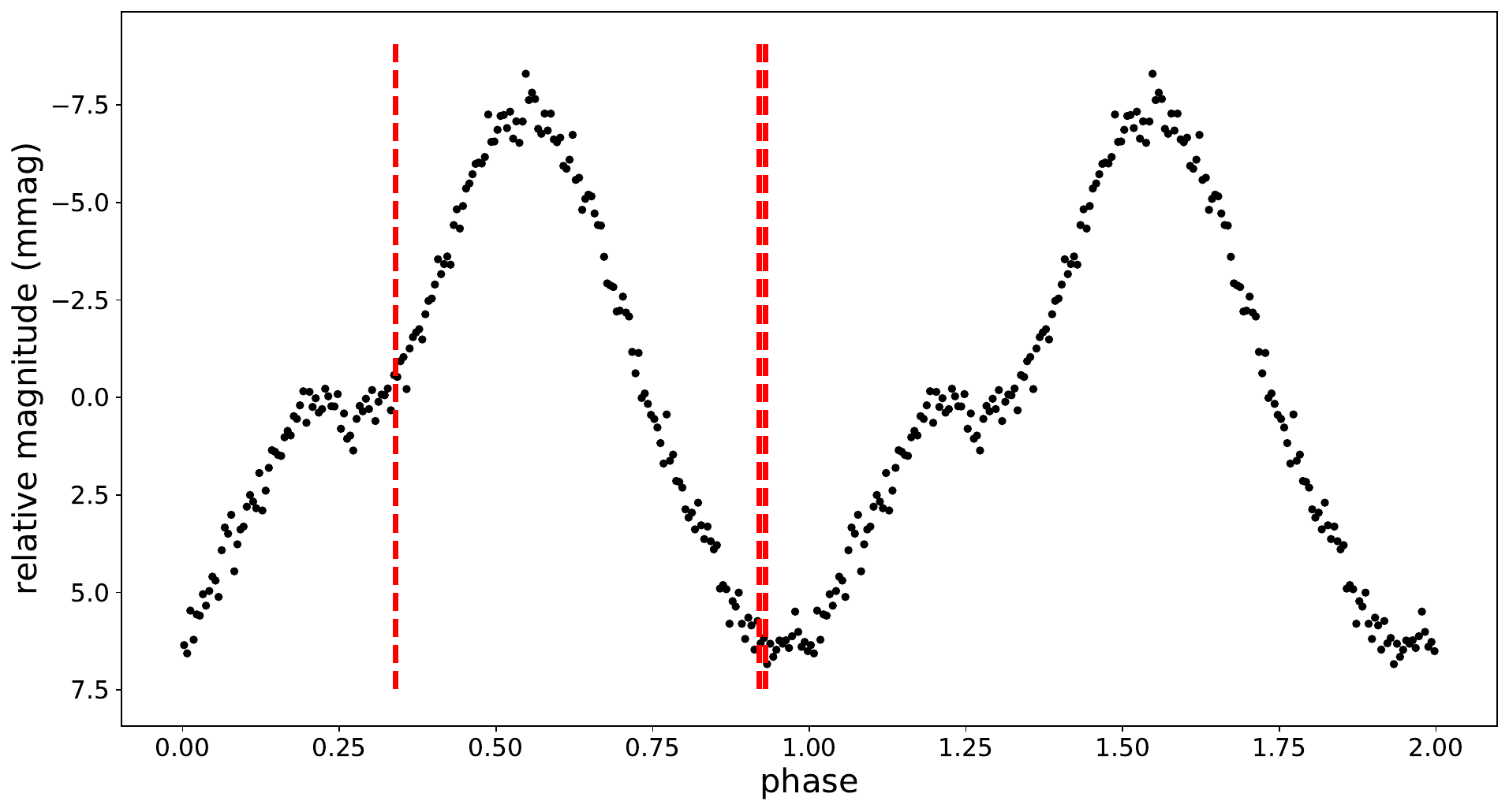}} 
\caption{ The phase folded light curve of KIC~10685175. The positions of three spectral observations are marked as dotted lines in the figure.
\label{fig:lc}} 
\end{figure}

For clarity, the phase folded light curve of KIC~10685175 is shown in Figure~\ref{fig:lc} and the rotation phases of three spectral observations are marked as dotted lines in the figure. The data, time zero-point, T$_0$=BJD~2,458,711.21391, and the rotation period, $P_{\rm rot} = 3.1028$\,d, are all from \cite{2020ApJ...901...15S}. 

\begin{table*}
  \centering
  \caption{Detailed information for the observations of KIC~10685175. The rotation phases are calculated based on the time zero-point T$_0$=BJD~2,458,711.21391 and the rotation period $P_{\rm rot} = 3.1028$\,d from \cite{2020ApJ...901...15S}.}
  \begin{tabular}{ccccccc}
    \hline
    HJD & Instrument & Resolution & SNR & Mode & Exposure & Rotation \\
     & & & at 6120\,$\rm \AA$ & & time (s) & phase\\
    \hline
    2459458.72644 & ESPaDOnS & 81\,000 & 60   &  unpolarized & 1300 & 0.92\\
    2459458.78562  & ESPaDOnS & 81\,000 & 60   &  unpolarized & 1300 & 0.93\\
    2459872.72369  & ESPaDOnS & 68\,000 &  90   &  circularly polarized & 1600$\times$4 & 0.34\\
  \hline
  \end{tabular}
    \label{tab:obs}
\end{table*}

%The data is reduced using Libre-ESpRIT \citep{1997MNRAS.291..658D}, which is the pipeline built for the ESPaDOnS. The reduction
%includes optimal spectrum extraction and normalization. {\red The radial velocity obtained by the pipeline, which is
%  about 3.5\,km~s$^{-1}$, is also corrected in the data process.}

%In order to improve the SNR, we first reduced the resolution of two unpolarized spectra to 47000 (green and blue spectra in Figure~\ref{fig:nosplit}), and then combined them into one (red solid spectrum in Figure~\ref{fig:nosplit}). In this way, SNR increased from 60 to 84. 

\section{Spectral analysis}
\label{Spectral analysis}

The effective temperature, $T_{\rm eff}$, and surface gravity, $\log g$, of KIC~10685175 have been measured in several works.
\citet{2018AA...616A...1G} derived $T_{\rm eff}$=7900\,K from {\it Gaia} DR2 Bp, Rp and G band photometry.  
\cite{2022AA...658A..91A} derived $T_{\rm eff}$=8000\,K and $\log g$=4.28 using {\it Gaia} DR3 photometry together with other multi-bands
of 2MASS and AllWISE. 
However, there are no stellar parameters based on high-resolution spectra for this star, and most measurements are from photometry and
isochrone fitting. As LAMOST provides low-resolution spectra for this star, some works, such as \cite{2022yCat.5156....0L},
derived $T_{\rm eff}$ and $\log g$ based on low-resolution spectra. The parameters for this star found in the literature are listed in Table~\ref{tab:spec_param}.

\begin{table*}
  \centering
  \caption{The parameters of KIC~10685175 in the literature. }
  \resizebox{\linewidth}{!}{
  \begin{tabular}{lllll}
    \hline
    $T_{\rm eff}$ (K) & $\log g$\,(cgs) & [Fe/H]\,(dex) & Method & Reference \\
    \hline
    7900$\pm$100 & - & - & three {\it Gaia} broad-band photometric measurements & \citet{2018AA...616A...1G} \\
  8000$\pm$280 & 4.28 & - & Gaia EDR3 + 2MASS + AllWISE photometry & \citet{2022AA...658A..91A} \\
  8000 & 4.04 & -0.11 & KIC griz photometry & \citet{2012ApJS..199...30P}\\
   8200$\pm$300 & 4.04 & -0.11 & KIC griz + 2MASS JHK + intermediate-bandD51 filter photometry & \citet{2014ApJS..211....2H}\\
  8000$\pm$30 & 4.08 & -0.03 & LAMOST low-resolution spectrum & \cite{2022yCat.5156....0L} \\
  7810$\pm$40 & 4.36 & -0.07 & LAMOST low-resolution spectrum & \cite{2022AA...662A..66X}\\
  8000 & 4.20 & -0.49 &  & this work\\
    \hline
  \hline
  \end{tabular}
  }
    \begin{tablenotes}
      \small
      \item Note: [Fe/H]=$\log_{10}(N_{\rm Fe}/N_{\rm H})-\log_{10}(N_{\rm Fe}/N_{\rm H})_\odot$, $\log_{10}{A({\rm Fe})}=\log_{10}(N_{\rm Fe}/N_{\rm H})+12$, where $A$(Fe) is the Fe abundance. In $\it iSpec$, $A({\rm Fe})_\odot$ is 7.5, following \cite{1998SSRv...85..161G}.
    \end{tablenotes}
  
    \label{tab:spec_param}
\end{table*}
 
Here, we used the unpolarized spectra to derive the stellar parameters including $T_{\rm eff}$, $\log g$, [Fe/H], [$\alpha$/Fe] and $V_{mic}$ for KIC~10685175. In this paper, the value of $\log g$ is in CGS system, and the unit of [Fe/H] and [$\alpha$/Fe] is dex.

The equivalent width (EW) measurements were performed using the program {\it iSpec} \citep{2014A&A...569A.111B, 2019MNRAS.486.2075B}. 
In the EW method, {\it iSpec} measures EWs of several selected lines by fitting them with Gaussian functions. The theoretical EWs can
be calculated considering the excitation equilibrium and ionization balance. A least-squares algorithm is applied in each iteration
and the optimized parameters are selected by minimizing the differences between the observed and theoretical EWs. 

The EW method requires unblended lines. We inspected \ion{Fe}{i} and \ion{Fe}{ii} lines one by one to make sure that they are unblended and their EWs are smaller than 100\,m$\rm \AA$. Finally, 12 lines including 11 \ion{Fe}{i} lines and one \ion{Fe}{ii} line were selected. 

{\it iSpec} will derive Fe abundances based on the EW of each Fe line. The Fe abundances are assumed to be independent with reduced equivalent width ($\log_{10} \frac{EW}{\lambda}$), the lower energy levels, the ionization states which is sensitive to micro-turbulance velocity, effective temperature, $\log g$, respectively.  

Since we only have one \ion{Fe}{ii} line, it is unreliable to derive $\log g$ based on the ionization balance. Here, we calculated $\log g$ from photometry. Assuming $T_{\rm eff}=8000$\,K, with parallax and magnitude from {\it Gaia} DR3, we have $\log g=4.4 \pm 0.1$.

We fixed $\log g=4.4$ and give initial values of $T_{\rm eff}$, [Fe/H], [$\alpha$/Fe] and $V_{mic}$. 
We started from [$T_{\rm eff}$, [Fe/H], [$\alpha$/Fe], $V_{mic}$]=[8000\,K, -0.1, 0.04, 2.5\,km~s$^{-1}$], where the initial values of $T_{\rm eff}$ and [Fe/H] are from the average values from Table~\ref{tab:spec_param}, [$\alpha$/Fe]=$-$0.4[Fe/H] was calculated using the relationship in \citet{2022arXiv221211639M}, and $V_{mic}$ was estimated using an empirical relation considering the effective temperature, surface gravity and metallicity which was constructed in $\it iSpec$.
Then these four parameters were gradually changed to finally make sure that there is no trend between [Fe/H] and lower excitation energy; there is no trend between [Fe/H] and reduced EWs; and the final [Fe/H] is the average value of those of 12 lines. 

Using the atomic line-list extracted from VALD, atmosphere model from ATLAS \citep{castelli2004}, and based on MOOG code \citep{Sneden2012},
we derived the final result [$T_{\rm eff}$, [Fe/H], [$\alpha$/Fe], $V_{mic}$]=[8250 $\pm$ 200\,K, -0.4 $\pm$ 0.2, 0.16 $\pm$ 0.1, 1.7$\pm$0.2\,km~s$^{-1}$]. 
The uncertainties of $T_{\rm eff}$ and $V_{mic}$ were decided from that the slopes -- $T_{\rm eff}$ and $V_{mic}$ the slopes of the trend of [Fe/H] v.s. lower excitation energy and [Fe/H] v.s. reduced equivalent width should be smaller than 0.1, the uncertainty of [Fe/H] was from standard deviation of those of 12 lines, and the uncertainty of [$\alpha$/Fe] was propagated from that of [Fe/H].
We also examined that the ionization equilibrium is satisfied using the fixed $\log g=4.4$.

$\rm{[Fe/H]}=-0.4$ is very different compared with the measurements in the literature, where the results have been derived from photometry
and evolution models that are less suitable for Ap stars. The low [Fe/H] also indicates that the Fe abundance of this star is lower than that
of an Ap star with $T_{\rm eff} =  8000$\,K, although it is typical for Ap stars much cooler than KIC\,10685175 (see fig.\,5 of
\citealt{2018MNRAS.480.2953G}.)

\begin{figure}[htb]
%\center{\includegraphics[width=13cm]  {compare.png}} 
\center{\includegraphics[width=8cm]  {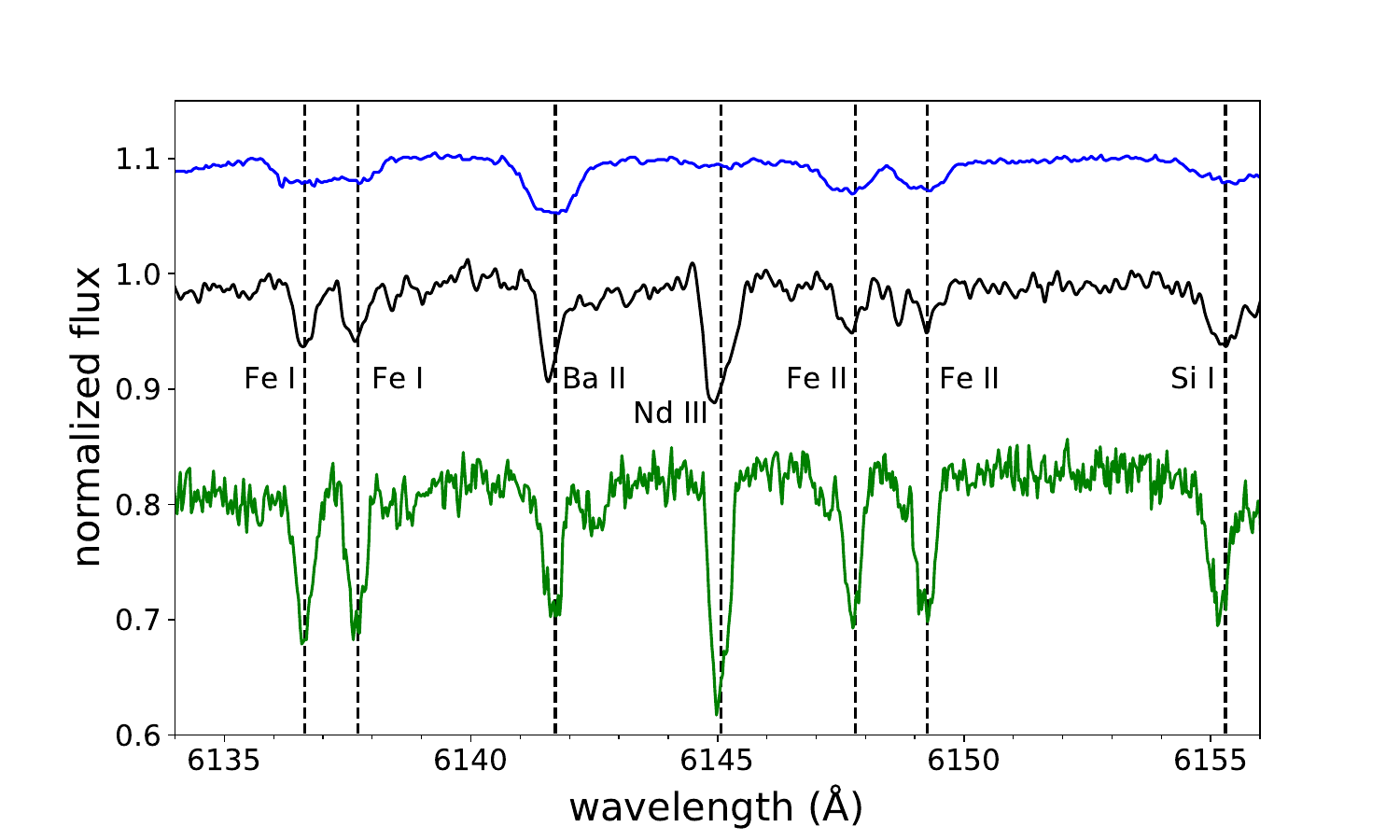}} 
\center{\includegraphics[width=8cm]  {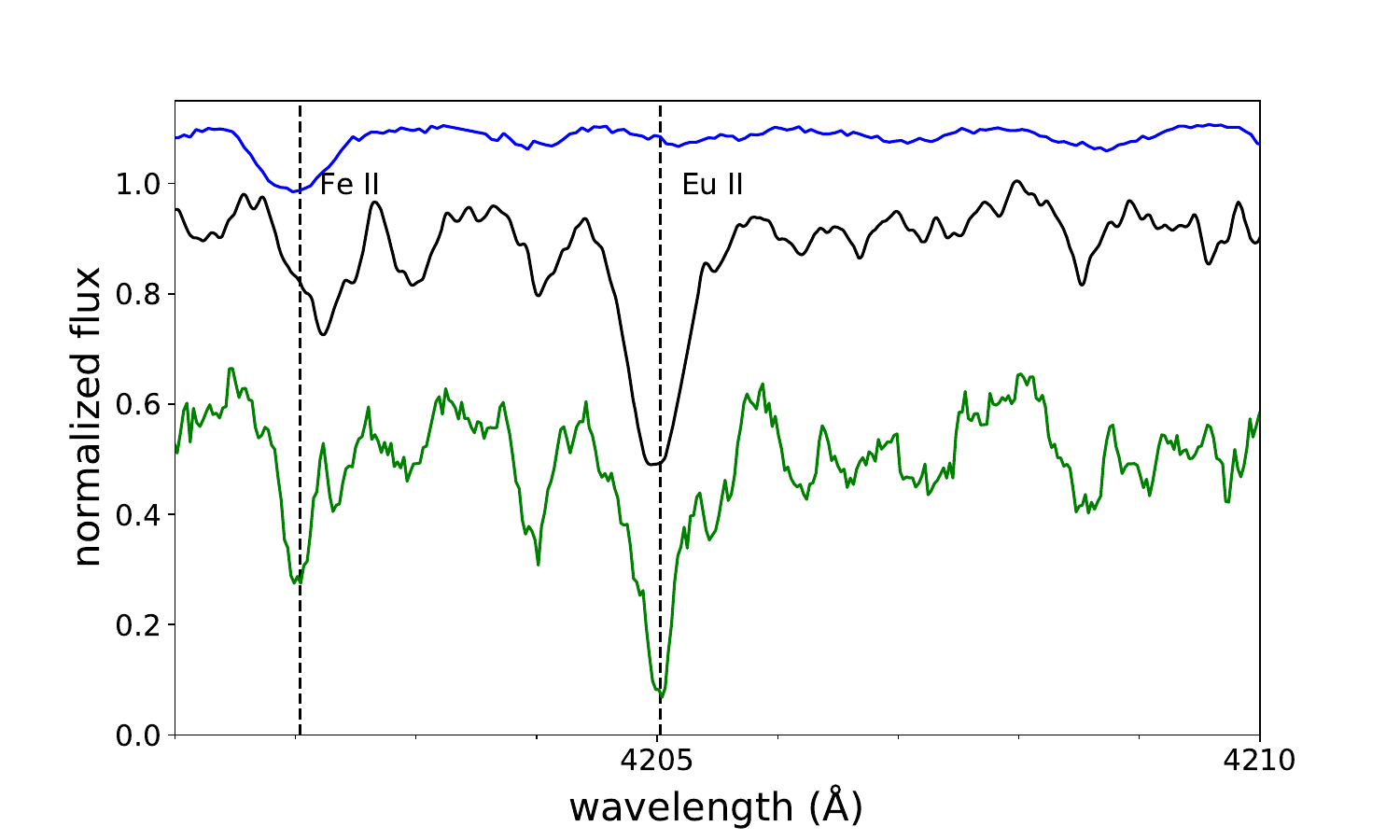}} 
\caption{Comparison of the spectra in the vicinity of the Nd and Eu lines between three A-type stars: the spectrum the star with normal
  abundance, HD~186689 (top -- blue),
  our target, KIC~10685175 (middle -- black), and the spectrum of the known roAp star, $\alpha$\,Cir (bottom -- green). Strong spectral lines
  are marked with black dashed lines. The abscissa is wavelength in ($\rm \AA$) and the ordinate is normalized intensity. For clarity, the spectra of HD~186689 and $\alpha$\,Cir are shifted.  
\label{fig:compare}} 
\end{figure}

In the following we compared the spectrum of KIC~10685175 with another Ap star and an A type star with normal chemical abundance.
The chemically normal star we chose is HD~186689, with fundamental parameters $T_{\rm eff} =7950 $\,K, $\log g =4.16$, and [Fe/H]=$-$0.33 \citep{2016A&A...595A...1G,2023A&A...674A...1G} which are similar to those of KIC~10685175. 
The $v\sin i$ for this star is 34\,km~s$^{-1}$\citep{2003A&A...398.1121E}, which is slightly larger than that of KIC~10685175. 
The spectrum was obtained from the ELODIE\footnote{http://atlas.obs-hp.fr/elodie/} database and
is shown in Fig.~\ref{fig:compare}. Compared with HD~186689, KIC~10685175 has slightly weaker Fe lines but Nd and Eu show significant overabundance, which is a typical feature for Ap stars.
We also compared the spectrum of KIC~10685175 with the spectrum of another Ap star with similar $T_{\rm eff}$, $\alpha$~Cir.
The $T_{\rm eff} = 7900$\,K, $\log g = 4.2$ and $v\sin i$=13.2\,km~s$^{-1}$ of $\alpha$ Cir \citep{2018MNRAS.480.2953G,2012A&A...542A.116A} are close to those of KIC~10685175.
The spectrum of $\alpha$ Cir was obtained by \cite{2020PASP..132j5001H} using the High Resolution Spectrograph (HRS) attached to the Southern
African Large Telescope (SALT). In Figure~\ref{fig:compare}, we show that for KIC~10685175, the Fe lines are much weaker than those
in $\alpha$ Cir. For Eu and Nd lines, by calculating equivalent widths of each line, we confirmed that two stars have similar intensities for \ion{Eu}{ii} line, and the \ion{Nd}{iii} line of KIC~10685175 is less enhanced.

\section{The magnetic field of KIC 10685175}
\subsection{The surface magnetic field}

High-resolution spectroscopic observations can be used to determine the mean magnetic field modulus,
$\langle B \rangle$ \citep{1989FCPh...13..143M}, by measuring the wavelength shifts between the magnetically split components using the relation:

\begin{equation}
\lambda_r-\lambda_b=g{k\lambda_0^2}\langle B \rangle\label{eq:b}.
\end{equation}

\noindent Here, $\lambda_r$, $\lambda_b$, and $\lambda_0$ represent the wavelength of the red and blue components, and the central wavelength, respectively;
$g$ is the Land\'e  factor, and $k=4.67{\times}10^{-13}$\,{\AA}$^{-1}{\rm G}^{-1}$. 

%Two magnetic sensitive Fe lines, Fe~II $\lambda$\,6147\,\AA{} and Fe~II $\lambda$\,6149\,\AA{}, are used (Land\'e  factor, $g$=2.7).
%Considering the spectral resolution of ESPaDOnS at this wavelength is about 81000 which means the minimum resolved wavelength is about 0.075\,\AA{}, according to eqn~\ref{eq:b}, a minimum magnetic field strength of 1.6\,kG can be observed by ESPaDOnS. 

\begin{figure}[htb]
\center{\includegraphics[width=8cm]  {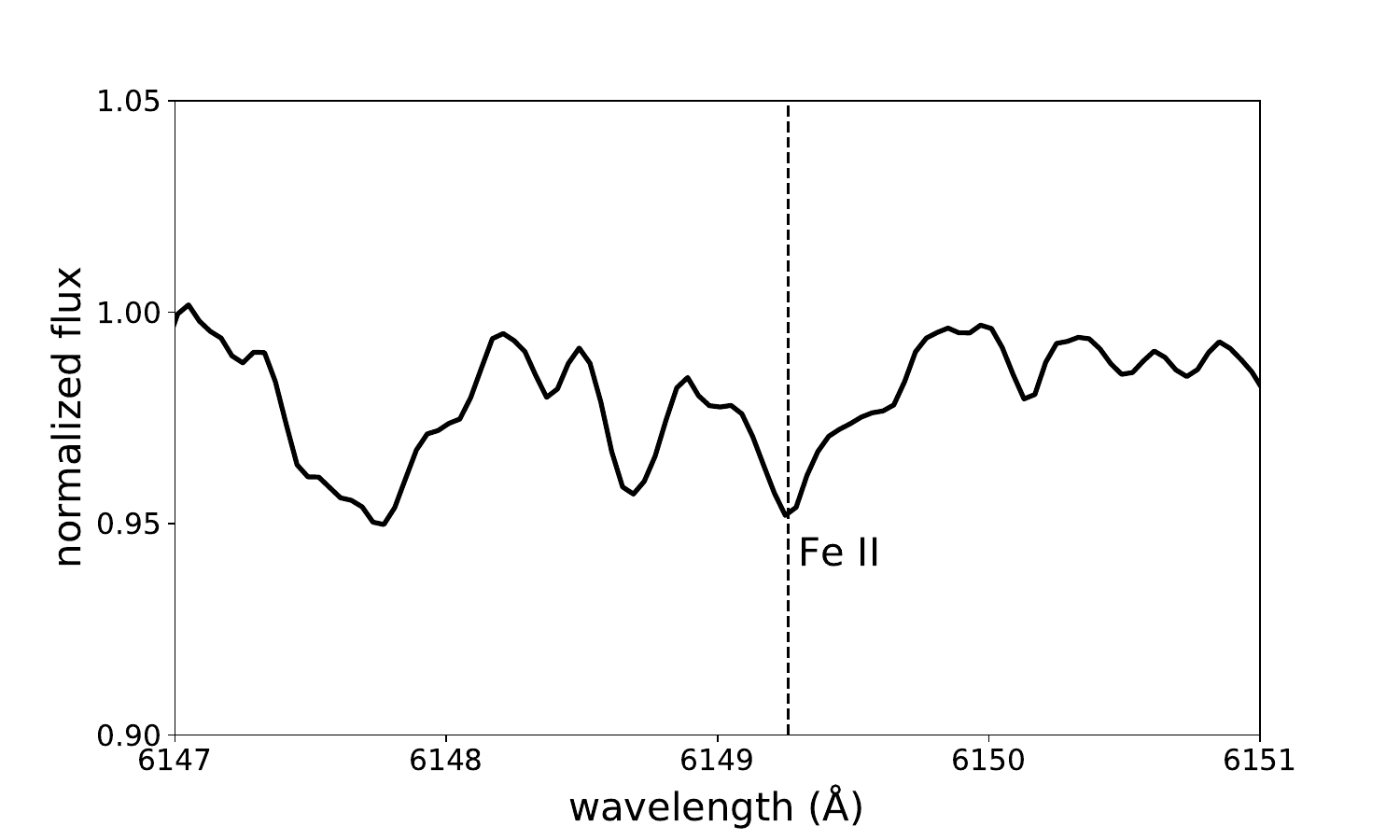}} 
\caption{The combined spectrum ($R = 81\,000$) of KIC~10685175 zoomed in the spectral region containing the magnetically sensitive line \ion{Fe}{ii}\,6149\,\AA{}.
\label{fig:nosplit}} 
\end{figure}

The widely used magnetically sensitive doublet \ion{Fe}{ii}\,6149\,\AA{} has been inspected for the presence of magnetically split lines. However, as we show in
Fig.~\ref{fig:nosplit}, this line does not exhibit magnetically split structure.
%For this star $v\sin i \approx 19$\,km~s$^{-1}$,
%which results in line broadening of Fe~II $\lambda$\,6149\,\AA{} of about 0.4\,$\rm \AA$. To observe the split components of this line, a surface magnetic field above 8\,kG would be needed. Thus it is reasonable that the split components are not detected.
%{\bf Comment: I suggest to skip the sentence with the expected 8kG - the problem is that with the field pole strength of 6kG you have to see some additional broadening, unless the magnetic field structure is more complex than a dipole.}
We also have inspected other magnetically sensitive lines which have large Land\'e factors ($\textsl{\textrm{g}}_{\rm eff}\ge$1.5) and
doublet patterns, such as \ion{Fe}{i} 6336.82\,\AA{} ($\textsl{\textrm{g}}_{\rm eff}=$2.0) and \ion{Cr}{ii} 5116.04\,\AA{} ($\textsl{\textrm{g}}_{\rm eff}=$2.92), but no magnetically split pattern are detected. The relatively high $v\sin i$ for this star significantly inhibits the detection of magnetically split lines. 
Besides, the profile variation due to the spots can also contaminate the line splitting or broadening from the Zeeman effect, which makes the line splitting less likely to be detected.

\subsection{The longitudinal magnetic field}

The reduced circularly polarized observations provide Stokes intensity $I$, circular polarization $V$ parameters, and diagnostic $N$ spectra. The mean longitudinal magnetic field, $\langle B_\ell \rangle$, is determined through:

\begin{equation}
\langle B_{\rm \ell} \rangle=-2.14\times10^{11}\frac{\int{v}V(v)dv}{{\lambda_0}g_\mathrm{eff}c\int[I_c-I(v)]dv}.
\label{eq:bl}
\end{equation}

\noindent Here, $v$ is the velocity offset from the line center, and the unit is km\,s$^{-1}$; $\lambda_0$ and $g_\mathrm{eff}$ are the effective wavelength
and the effective Land\'e factor used for normalization, respectively \citep{1979A&A....74....1R,1997MNRAS.291..658D}.

The noise level is greatly reduced with the help of the Least Square Deconvolution technique \citep[LSD;][]{1997MNRAS.291..658D,2018MNRAS.474.4956F}. 
In this technique it is assumed that each spectral line can be described by the same profile with a different scale factor which depends on
the central wavelength, the line strength, and the magnetic sensitivity.
%Then the mean profiles of intensity ($I$) and the circular
%polarization parameter ($V$) of several lines can be calculated from their observed line profiles and the scale factors.

The line list was constructed using the Vienna Atomic Line Database \citep[VALD;][]{Ryabchikova_2015} assuming $T_{\rm eff}=8200$\,K and $\log g=4.4$.
All the lines were inspected to make sure that they are deeper than 5$\%$ of the continuum and not significantly blended with other lines.
Lines contaminated by telluric absorption or located in hydrogen line wings were removed from the line list. The information about the selected line list 
is presented in Table~\ref{tab:obs2}. 

\begin{table*}
  \centering
  \caption{Numbers of the lines in line lists and the mean longitudinal magnetic field calculated from different elements.}
  \label{tab:obs2}
  \begin{tabular}{llllllll}%{p{3cm}p{3cm}p{5cm}}
    \hline\hline\noalign{\smallskip}
    Element & Number   & The mean  & The mean  & The mean & The mean longitudinal  & FAP & Detectable\\
     		 & of lines   &	depth       & Land\'e factor & wavelength ($\rm \AA$) &  magnetic field (G) &  & \\	
    \hline\noalign{\smallskip}
    all & 208 & 0.49 & 1.26 & 4914 & $-226\pm39$ & $1.7\times10^{-6}$  & DD\\
    \ion{Ca}{i} & 28 & 0.51	& 1.18  &  5358 &  $-115\pm57$ & 0.011 & ND\\
    \ion{Cr}{ii} & 20 & 0.46	&1.20  &  4718 & $-398\pm105$ & $8\times10^{-4}$ & MD\\
    \ion{Fe}{i} & 75 & 0.49	& 1.33  &  5009 & $-248\pm47$ &  $6\times10^{-6}$ & DD\\
    \ion{Fe}{ii} & 21 & 0.61	&1.14  &  4798 & $-254\pm55$ & $6\times10^{-4}$ & MD\\
  \hline
  \end{tabular}
\end{table*}

\begin{figure}[htb]
%\center{\includegraphics[width=13cm]  {compare.png}} 
\center{\includegraphics[width=8cm]  {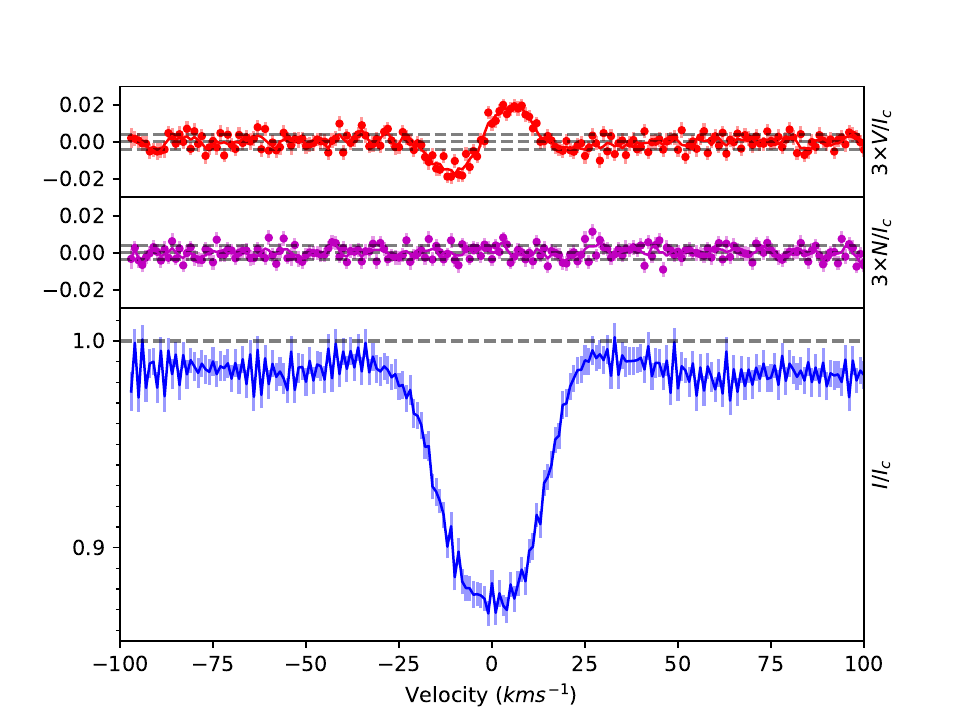}} 
\caption{The profile of Stokes $V$, diagnostic null, and $I$ (from top to bottom) calculated for KIC~10685175.
 %The line list is built through VALD. The velocity range used to
% calculate LSD profiles and magnetic fields is -30 to 30\,km~s$^{-1}$. 
\label{fig:lsd}} 
\end{figure}

In Fig.~\ref{fig:lsd}, we show for the line list constructed for all elements the LSD profiles of Stokes~$I$, $V$, and diagnostic null.
LSD profiles were calculated in the velocity range of -30 to 30\,km~s$^{-1}$ using a step of 1\,km~s$^{-1}$.
The normalized Land\'e factor is 1.26, and the normalized wavelength of 4914\,$\rm \AA$. 
We followed the generally adopted procedure to use the false alarm probability (FAP) based on reduced $\chi^2$ test statistics \citep{1992A&A...265..669D,1997MNRAS.291..658D}: the presence of a Zeeman signature is considered as a definite detection (DD) if FAP $\leq$ 10$^{-5}$ ,
  as a marginal detection (MD) if 10$^{-5}$ $\textless$ FAP $\leq$ 10$^{-3}$, and as a non-detection (ND) if FAP $\textgreater$ 10$^{-3}$.
 %Due to the presence of chemical spots on the stellar surface, the line profile changes, which may cause the Stokes~$I$ profile not to be centered at 0~km/s exactly.
 % {\bf Comment: skip the sentence of the shift of the Stokes I profile - if you compare the position of this profile for all line lists, then  you will see that there is no shift due to the spots. The tiny shift of the position is due to the radial velocity, not due to spots.}
  The $V$ spectrum exhibits a clear Zeeman feature indicating the presence of a magnetic field.  Using Eq.~\ref{eq:bl}, the mean longitudinal magnetic field
  is calculated to be $\langle B_{\rm \ell} \rangle = -226\pm39$\,G. The FAP is listed in Table~\ref{tab:obs2}.
  %In Section~\ref{discussion} we will compare it with the predicted polar magnetic field.

\subsection{The comparison of the measured and the predicted magnetic field strengths}

In the recent study of \cite{2020ApJ...901...15S} it was suggested that  KIC~10685175 possess a strong magnetic field with the dipole magnetic field
strength of 6\,kG.  The theoretical pulsation model of \cite{2005MNRAS.360.1022S} was used to
compare the observations of KIC~10685175 with theoretical predictions.
Following \cite{1950MNRAS.110..395S}, and assuming a dipole magnetic field inclined by an angle $\beta$ to the rotation axis, a polar magnetic field
can be estimated through:

\begin{equation}
    B_\ell=\frac{1}{20}\frac{15+u}{3-u}B_p(\cos\beta\cos i +\sin\beta\sin i \cos (2\pi\Phi)).
\label{eq_b}
\end{equation} 

\noindent 

The limb darkening coefficient $u=0.5$ was extracted from \cite{2011A&A...529A..75C}, using $\log g=4.5$, $T_{\rm eff}=8000$\,K and metallicity $Z=0$.
The value $Z$ was derived through $Z$=10$^{[\rm {Fe}/\rm H]}$$\times$Z$_\odot$, where [Fe/H]=$-$0.4 and Z$_\odot$=0.02.
Using the mean longitudinal magnetic field strength $B_\ell=-226 \pm 39$\,G determined in the rotational phase $\Phi=0.34$,
the stellar inclination and the magnetic obliquity, $(i, \beta)=(60^\circ, 60^\circ)$, from the pulsation model presented in \cite{2020ApJ...901...15S},
we estimated the polar magnetic field to be $4.8 \pm 0.8$\,kG. This is compatible with the theoretically predicted polar magnetic field of 6\,kG within 3$\sigma$ supporting the theoretical model of \cite{2005MNRAS.360.1022S}.

\section{Discussion and conclusion}
\label{discussion}

Using two high-resolution spectroscopic spectra and one circularly polarized spectrum we have studied the stellar parameters of the roAp star KIC~10685175 and measured its mean longitudinal magnetic field strength.
Compared to chemically normal stars, KIC~10685175 exhibits chemical peculiarities such as overabundance of Eu and Nd which is typical of
magnetic Ap stars. Fe lines however appear weaker than in other Ap stars with similar fundamental parameters.

\begin{figure*}[b]
  \centering
    \begin{subfigure}%{0.45\textwidth}
      \centering  
      \includegraphics[width=0.35\linewidth]{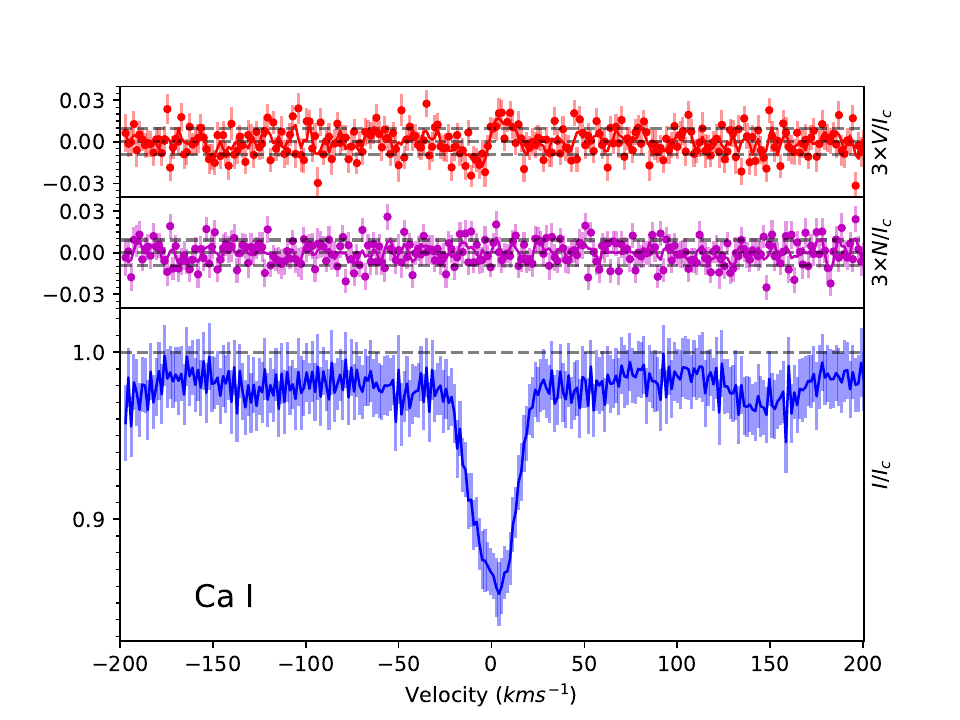}
        %\caption{caption\_for\_sub1}
        %\label{fig:sub1}
    \end{subfigure} 
\begin{subfigure}%{0.45\textwidth}
      \centering   
      \includegraphics[width=0.35\linewidth]{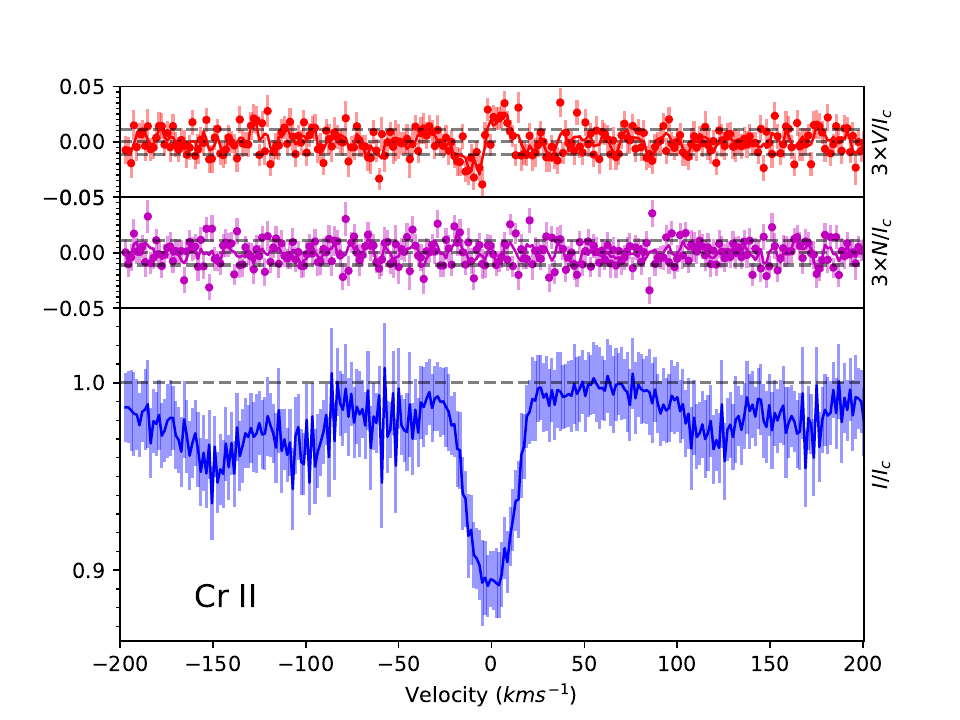}
        %\caption{caption\_for\_sub2}
        %\label{fig:sub2}
    \end{subfigure}
\begin{subfigure}%{0.45\textwidth}
      \centering   
     % \adjustbox{raise=0cm}{ 
      \includegraphics[width=0.35\linewidth]{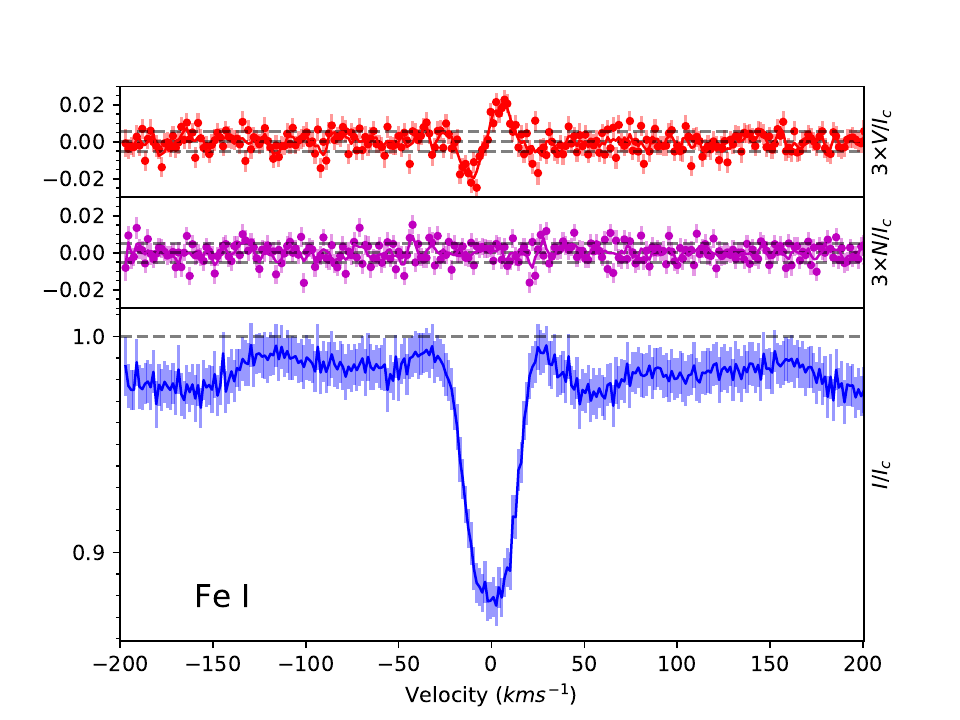}%}
        %\caption{caption\_for\_sub2}
        %\label{fig:sub2}
    \end{subfigure}
\begin{subfigure}%{0.45\textwidth}
      \centering   
      %\adjustbox{raise=0cm}{ 
      \includegraphics[width=0.35\linewidth]{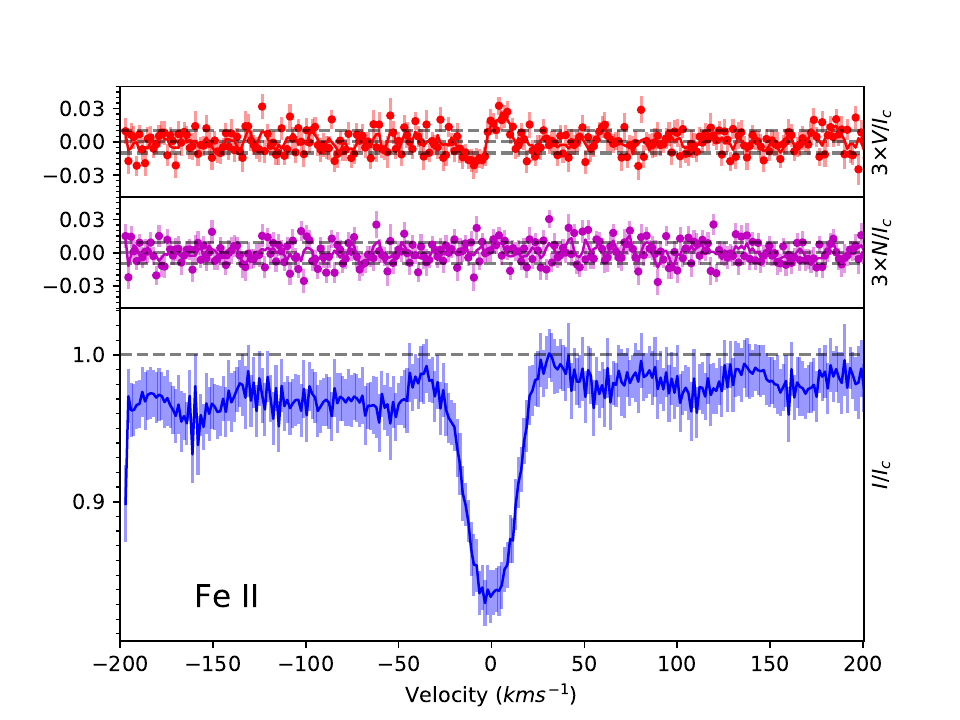}%}
        %\label{fig:sub2}
    \end{subfigure}
    \captionsetup{skip=11cm} 
    \captionsetup{position=top}  
\caption{
Same as Fig.~\ref{fig:lsd} but for different elements. 
\label{fig:lsd_ele}} 
\end{figure*}

Due to the lower number of available spectra, it is difficult to conclude whether chemical spots are present on the stellar surface.
An inhomogeneous surface element distribution can be explored by considering the differences in magnetic field strengths obtained using line lists
constructed for different elements. 
%Although the line profiles cannot give any clue for the element distribution, we can explore the elements distribution on the stellar surface through the mean longitudinal magnetic field. 
Assuming a dipolar magnetic field configuration, if a stronger mean longitudinal magnetic field is measured in the lines of a specific element,
we can conclude that this element tends to form a spot closer to the magnetic pole.
The results of the LSD technique applied to the line masks for \ion{Ca}{i}, \ion{Cr}{ii}, \ion{Fe}{i} and \ion{Fe}{ii} are displayed in Fig.~\ref{fig:lsd_ele}
and the results of our measurements of the mean longitudinal magnetic field are presented in Table~\ref{tab:obs2}.
Unfortunately, due to the low number of unblended lines
belonging to the rare earth elements, no corresponding line lists have been constructed. \cite{H_mmerich_2018} classified KIC~10685175 as A4V\,Eu. However,
there are too few Eu lines visible in the spectrum to construct and Eu line mask.

As we show in Table~\ref{tab:obs2}, only the mean longitudinal magnetic field using \ion{Fe}{i} lines is definitely detected and the 
result, B$_\ell$(\ion{Fe}{i})=$-248\pm47$\,G, is very close to the measurements obtained using all lines. 
This is reasonable because the number of \ion{Fe}{i} lines dominates the line list.
For \ion{Ca}{i}, we obtained B$_\ell$(\ion{Ca}{i})=$-115\pm57$\,G, but the FAP indicates no detection.
The magnetic fields for \ion{Fe}{ii} ($-254\pm55$\,G) and \ion{Cr}{ii} ($-398\pm105$\,G) are only marginally detected. 
The difference between the magnetic field strengths obtained for \ion{Cr}{ii}, \ion{Fe}{i} and \ion{Fe}{ii} is larger than 1$\sigma$. It may probably be due to
the inhomogeneous distribution of these elements, but is not significant enough to make a solid conclusion.
%, moreover, the dipolar configuration is just a rough assumption, it is hard to conclude that Cr and Fe distribute at different regions on the stellar surface.
%To study the inhomogeneous element distribution, additional observations covering more rotation phases are needed.}

\begin{figure}[htb]
\center{\includegraphics[scale=0.35]  {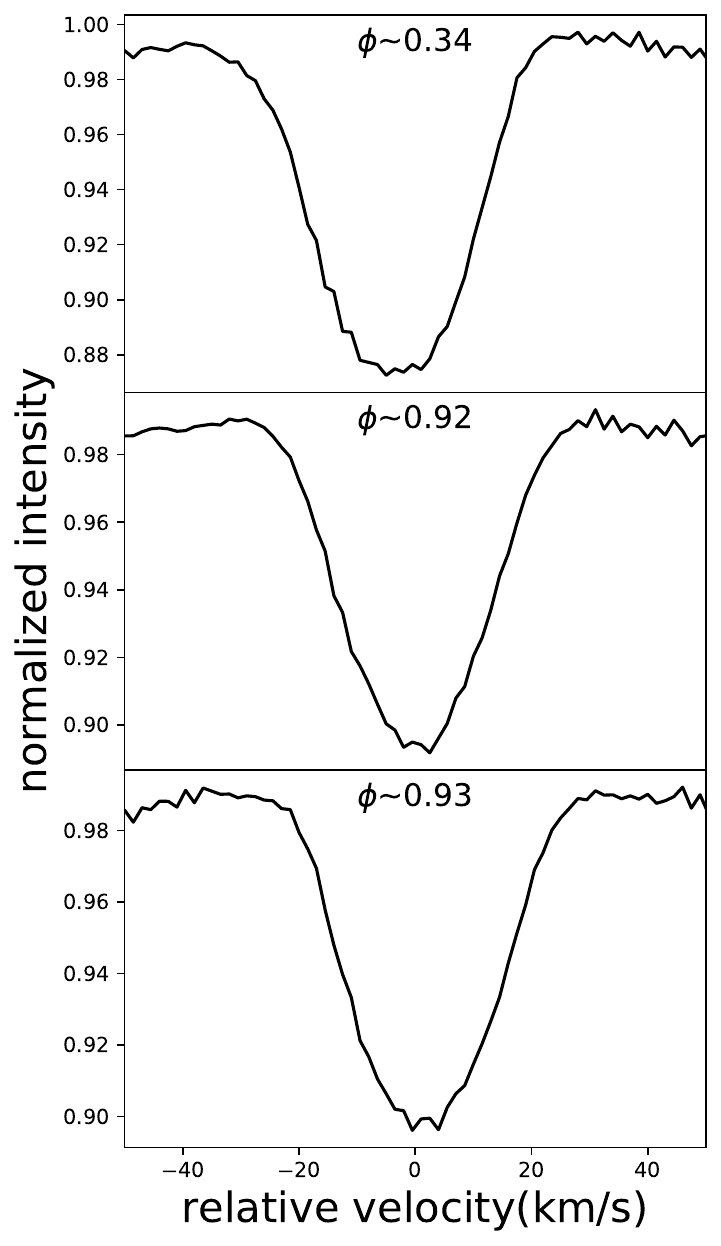}}
\caption{The LSD Stokes $I$ profiles calculated for all three available observations using all lines.
Rotation phase are marked on the top of each panel.
\label{fig:var_all}} 
\end{figure}

\begin{figure*}[b]
  \centering
    \begin{subfigure}%{0.45\textwidth}
      \centering   
      \includegraphics[width=0.18\linewidth]{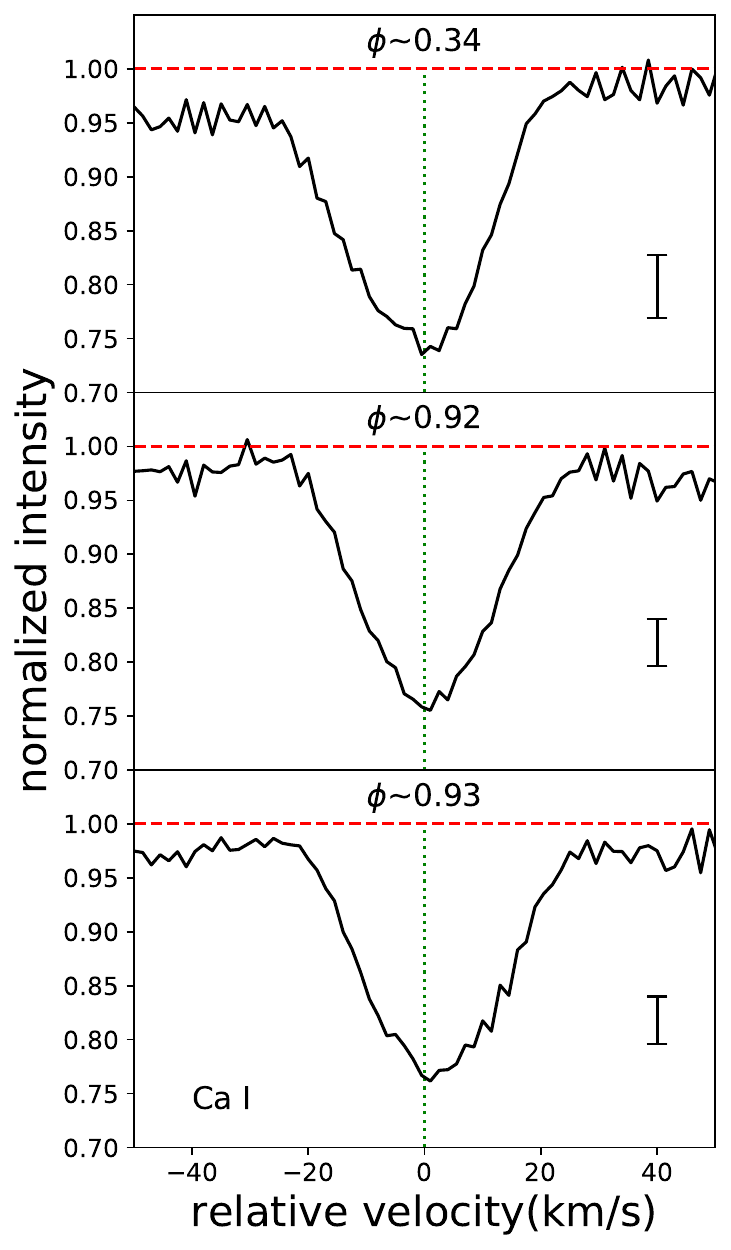}
    \end{subfigure} 
\begin{subfigure}%{0.45\textwidth}
      \centering   
      \includegraphics[width=0.18\linewidth]{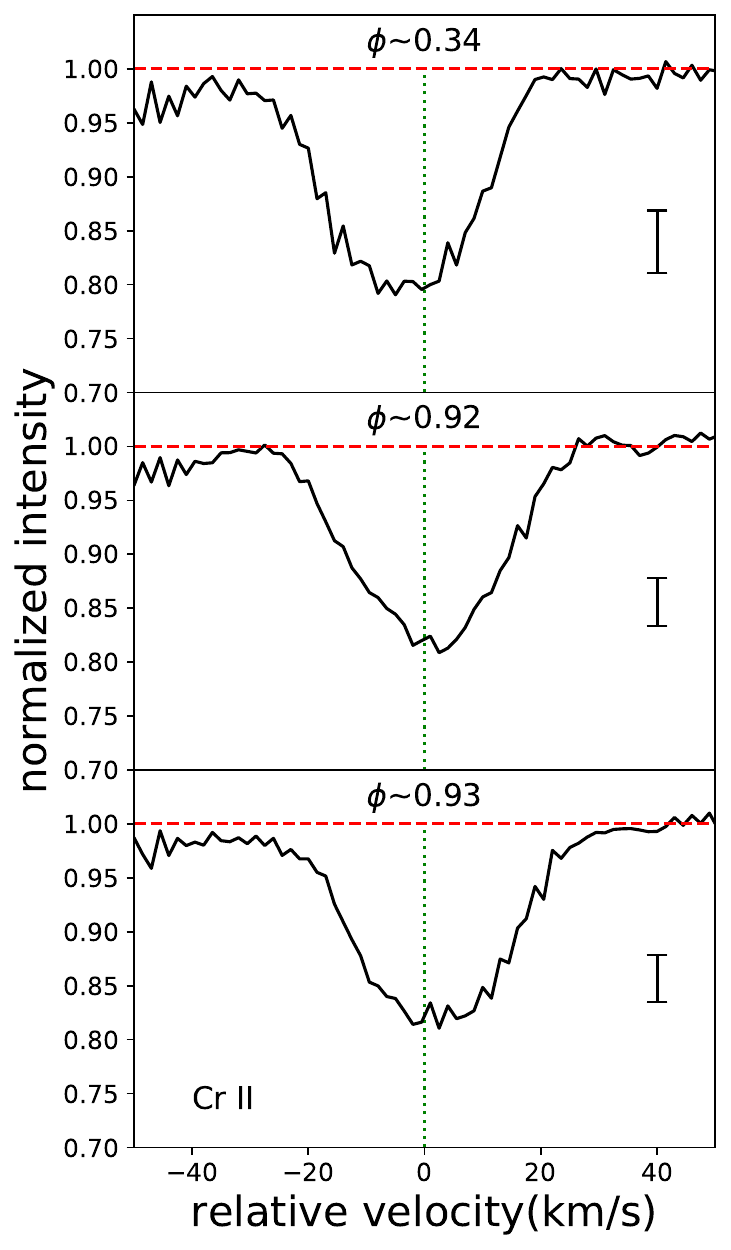}
    \end{subfigure}
\begin{subfigure}%{0.45\textwidth}
      \centering   
      \includegraphics[width=0.18\linewidth]{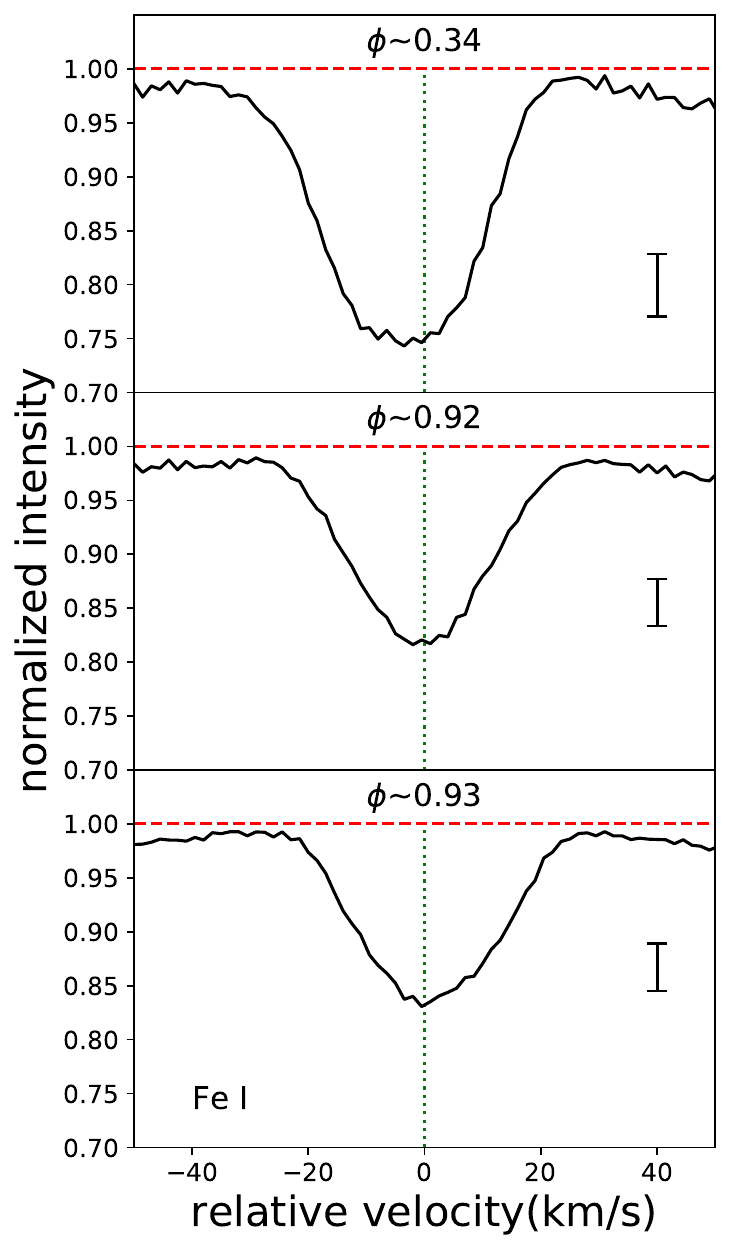}
    \end{subfigure}
\begin{subfigure}%{0.45\textwidth}
      \centering   
      \includegraphics[width=0.18\linewidth]{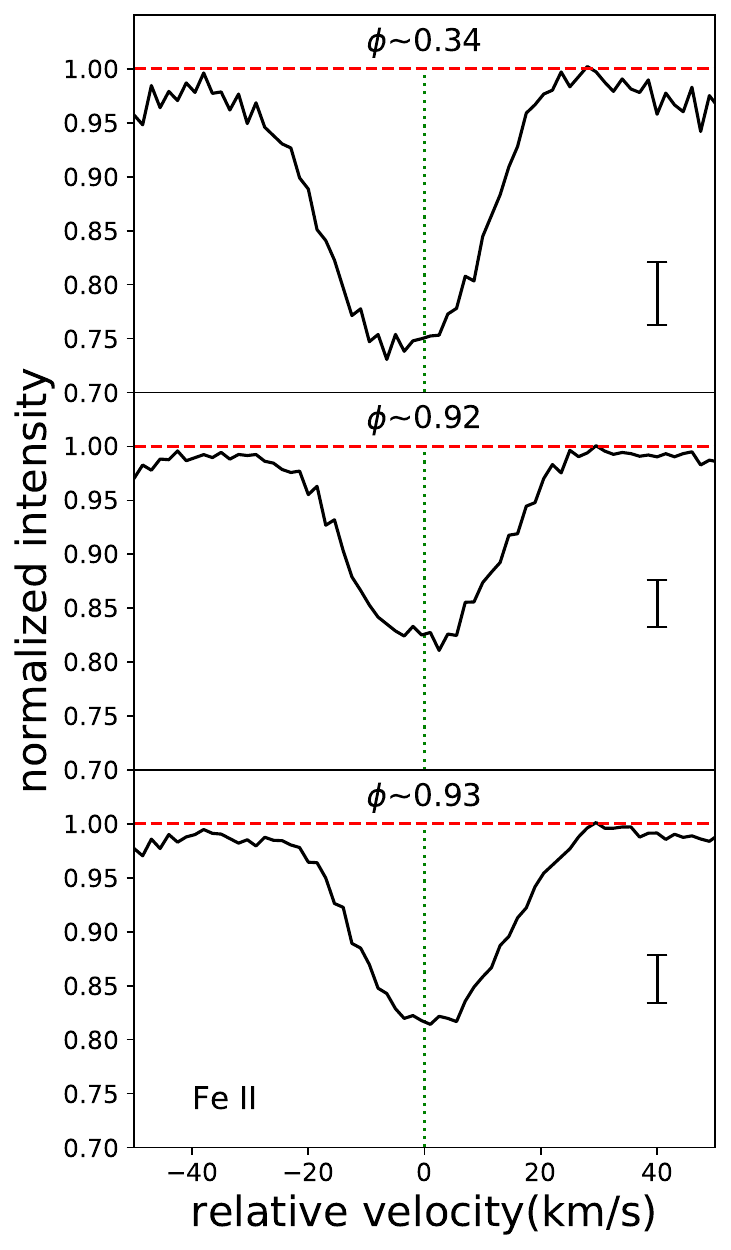}
    \end{subfigure}
\begin{subfigure}%{0.45\textwidth}
      \centering   
      \includegraphics[width=0.18\linewidth]{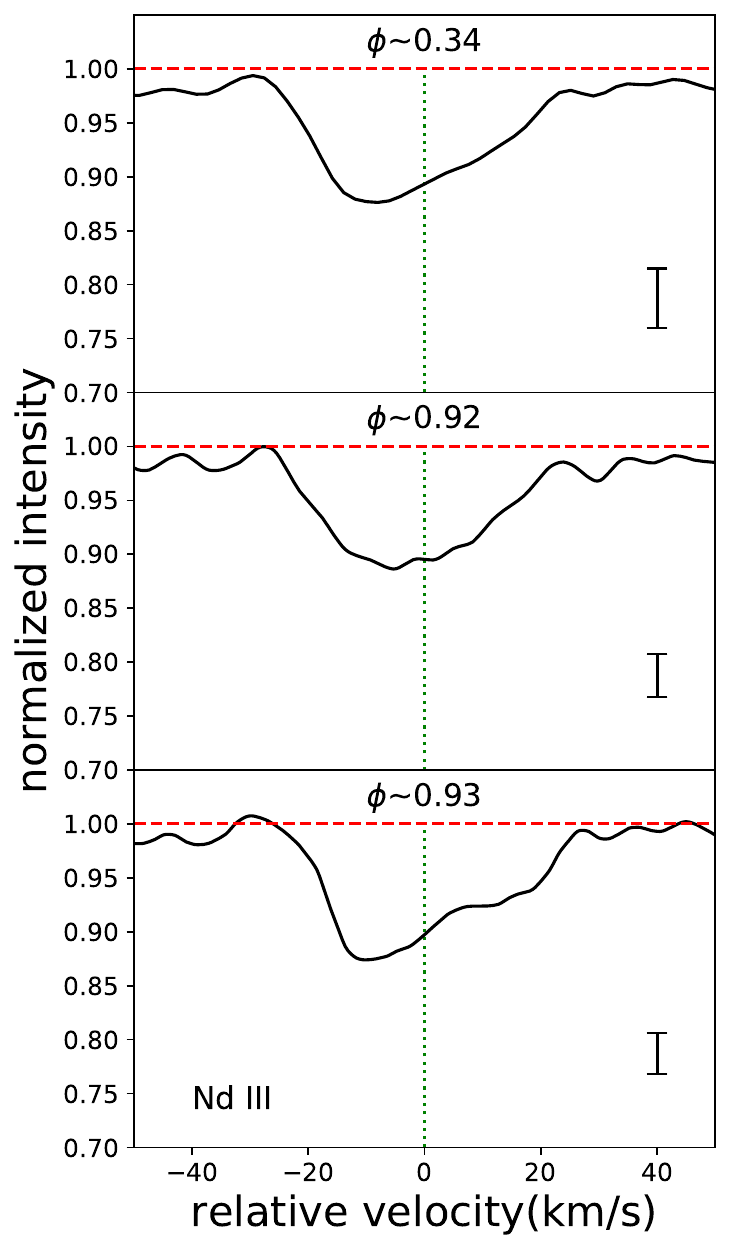}
    \end{subfigure}
    \captionsetup{skip=4cm} 
    \captionsetup{position=top} 
\caption{
Same as Fig.~\ref{fig:var_all} but for different elements. The rotation phases are marked at the top of each panel. For \ion{Nd}{iii}, the intensity profiles of 6145.07~$\rm \AA$ are directly compared, instead of calculating LSD profiles. 
\label{fig:var_ele}} 
\end{figure*}

Chemical spots on the surface of Ap stars usually cause significant variability of the spectral line profiles over the rotation period \citep{1958ApJS....3..141B}.
%(e.g. see \citealt{2001A&A...374..615K,2007MNRAS.376..651K}).
To further test the possible presence of chemical spots on the surface of KIC~10685175,
the LSD technique has been applied for all three available observations
to calculate the Stokes $I$ profiles.
Using the time zero-point T$_0$=BJD~2,458,711.21391 and the rotation period $P_{\rm rot} = 3.1028$\,d from \cite{2020ApJ...901...15S}, both unpolarized
spectra correspond to the rotation phases 0.92 and 0.93, respectively, and the polarized spectrum to the phase 0.34. 
For all three different rotation phases the Stokes $I$ profiles calculated for the line list containing all lines and those for the line lists constructed for different elements are presented in
Fig.~\ref{fig:var_all} and Fig.~\ref{fig:var_ele}. 
We also checked the Stokes $I$ profiles for Nd and Pr, but only \ion{Nd}{iii} 6145\,{\rm \AA} can be detected in all spectra. So in Fig.~\ref{fig:var_ele}, we compared the profiles of \ion{Nd}{iii} 6145\,{\rm \AA} directly without calculating LSD profiles.
No significant changes in the LSD Stokes $I$ profiles are detected, although the profiles
for \ion{Fe}{i} calculated for the phase 0.34 show slightly flat-bottom profiles compared to those observed in other phases. As the rotation phase coverage of our
data is rather poor, additional observations are needed to permit to conclude of the surface inhomogeneous element distribution.

Despite the fact that magnetically split lines were not detected in our spectra, under the assumption of the dipolar topology of the global magnetic field,
the strength of the measured mean longitudinal magnetic field,
$\langle B_\ell \rangle=-226\pm39$\,G in combination with already known stellar inclination and the magnetic obliquity suggest the polar magnetic field
to be $4.8 \pm 0.8$\,kG. This polar field strength is consistent with field strength predicted by the theoretical model of \cite{2005MNRAS.360.1022S}.

\section*{Acknowledgements}
This work was funded by the National Natural Science Foundation
of China (NSFC Grant No.12090040, 12090042, 12090044, U1731108 and 11833002) and the National Key R$\&$D Program of China (No.~2019YFA0405500). DWK acknowledges support from the Funda\c c\~ao para a Ci\^encia e a Tecnologia (FCT) through national funds (2022.03993.PTDC). 

This work is based on data collected at the Canada-France-Hawaii Telescope, and also uses data from the VALD database, European Space Agency mission {\it Gaia} (http://www.cosmos.esa.int/{\it Gaia}), and the SIMBAD database. The observation time is obtained through the Telescope Access Program (TAP).

Some of the observations in this paper were obtained with the Southern African Large Telescope (SALT) under proposal code 2017-1-SCI-023 (PI: Holdsworth).

\bibliographystyle{aa}
\bibliography{KIC10685175_ver17}{}

\begin{thebibliography}{47}
\expandafter\ifx\csname natexlab\endcsname\relax\def\natexlab#1{#1}\fi

\bibitem[{{Ammler-von Eiff} \& {Reiners}(2012)}]{2012A&A...542A.116A}
{Ammler-von Eiff}, M. \& {Reiners}, A. 2012, \aap, 542, A116

\bibitem[{{Anders} {et~al.}(2022){Anders}, {Khalatyan}, {Queiroz}, {Chiappini},
  {Ard{\`e}vol}, {Casamiquela}, {Figueras}, {Jim{\'e}nez-Arranz}, {Jordi},
  {Mongui{\'o}}, {Romero-G{\'o}mez}, {Altamirano}, {Antoja}, {Assaad},
  {Cantat-Gaudin}, {Castro-Ginard}, {Enke}, {Girardi}, {Guiglion}, {Khan},
  {Luri}, {Miglio}, {Minchev}, {Ramos}, {Santiago}, \&
  {Steinmetz}}]{2022AA...658A..91A}
{Anders}, F., {Khalatyan}, A., {Queiroz}, A.~B.~A., {et~al.} 2022, \aap, 658,
  A91

\bibitem[{{Babcock}(1958)}]{1958ApJS....3..141B}
{Babcock}, H.~W. 1958, \apjs, 3, 141

\bibitem[{{Balmforth} {et~al.}(2001){Balmforth}, {Cunha}, {Dolez}, {Gough}, \&
  {Vauclair}}]{2001MNRAS.323..362B}
{Balmforth}, N.~J., {Cunha}, M.~S., {Dolez}, N., {Gough}, D.~O., \& {Vauclair},
  S. 2001, \mnras, 323, 362

\bibitem[{{Bigot} \& {Kurtz}(2011)}]{2011A&A...536A..73B}
{Bigot}, L. \& {Kurtz}, D.~W. 2011, \aap, 536, A73

\bibitem[{{Blanco-Cuaresma}(2019)}]{2019MNRAS.486.2075B}
{Blanco-Cuaresma}, S. 2019, \mnras, 486, 2075

\bibitem[{{Blanco-Cuaresma} {et~al.}(2014){Blanco-Cuaresma}, {Soubiran},
  {Heiter}, \& {Jofr{\'e}}}]{2014A&A...569A.111B}
{Blanco-Cuaresma}, S., {Soubiran}, C., {Heiter}, U., \& {Jofr{\'e}}, P. 2014,
  \aap, 569, A111

\bibitem[{{Borucki} {et~al.}(2010){Borucki}, {Koch}, {Basri}, {Batalha},
  {Brown}, {Caldwell}, {Caldwell}, {Christensen-Dalsgaard}, {Cochran},
  {DeVore}, {Dunham}, {Dupree}, {Gautier}, {Geary}, {Gilliland}, {Gould},
  {Howell}, {Jenkins}, {Kondo}, {Latham}, {Marcy}, {Meibom}, {Kjeldsen},
  {Lissauer}, {Monet}, {Morrison}, {Sasselov}, {Tarter}, {Boss}, {Brownlee},
  {Owen}, {Buzasi}, {Charbonneau}, {Doyle}, {Fortney}, {Ford}, {Holman},
  {Seager}, {Steffen}, {Welsh}, {Rowe}, {Anderson}, {Buchhave}, {Ciardi},
  {Walkowicz}, {Sherry}, {Horch}, {Isaacson}, {Everett}, {Fischer}, {Torres},
  {Johnson}, {Endl}, {MacQueen}, {Bryson}, {Dotson}, {Haas}, {Kolodziejczak},
  {Van Cleve}, {Chandrasekaran}, {Twicken}, {Quintana}, {Clarke}, {Allen},
  {Li}, {Wu}, {Tenenbaum}, {Verner}, {Bruhweiler}, {Barnes}, \&
  {Prsa}}]{2010Sci...327..977B}
{Borucki}, W.~J., {Koch}, D., {Basri}, G., {et~al.} 2010, Science, 327, 977

\bibitem[{Castelli \& Kurucz(2004)}]{castelli2004}
Castelli, F. \& Kurucz, R.~L. 2004, New Grids of ATLAS9 Model Atmospheres

\bibitem[{{Claret} \& {Bloemen}(2011)}]{2011A&A...529A..75C}
{Claret}, A. \& {Bloemen}, S. 2011, \aap, 529, A75

\bibitem[{{Cui} {et~al.}(2012){Cui}, {Zhao}, {Chu}, {Li}, {Li}, {Zhang}, {Su},
  {Yao}, {Wang}, {Xing}, {Li}, {Zhu}, {Wang}, {Gu}, {Luo}, {Xu}, {Zhang},
  {Liu}, {Zhang}, {Yang}, {Cao}, {Chen}, {Chen}, {Chen}, {Chen}, {Chu}, {Feng},
  {Gong}, {Hou}, {Hu}, {Hu}, {Hu}, {Jia}, {Jiang}, {Jiang}, {Jiang}, {Jin},
  {Li}, {Li}, {Li}, {Liu}, {Liu}, {Lu}, {Mao}, {Men}, {Qi}, {Qi}, {Shi},
  {Tang}, {Tao}, {Wang}, {Wang}, {Wang}, {Wang}, {Wang}, {Wang}, {Wang},
  {Wang}, {Wang}, {Wang}, {Wang}, {Wang}, {Xu}, {Xu}, {Yang}, {Yu}, {Yuan},
  {Yuan}, {Zhai}, {Zhang}, {Zhang}, {Zhang}, {Zhao}, {Zhou}, {Zhou}, {Zhu}, \&
  {Zou}}]{2012RAA....12.1197C}
{Cui}, X.-Q., {Zhao}, Y.-H., {Chu}, Y.-Q., {et~al.} 2012, Research in Astronomy
  and Astrophysics, 12, 1197

\bibitem[{{Cunha}(2006)}]{2006MNRAS.365..153C}
{Cunha}, M.~S. 2006, \mnras, 365, 153

\bibitem[{{Cunha} \& {Gough}(2000)}]{2000MNRAS.319.1020C}
{Cunha}, M.~S. \& {Gough}, D. 2000, \mnras, 319, 1020

\bibitem[{{Donati} {et~al.}(2006){Donati}, {Catala}, {Landstreet}, \&
  {Petit}}]{2006ASPC..358..362D}
{Donati}, J.~F., {Catala}, C., {Landstreet}, J.~D., \& {Petit}, P. 2006, in
  Astronomical Society of the Pacific Conference Series, Vol. 358, Solar
  Polarization 4, ed. R.~{Casini} \& B.~W. {Lites}, 362

\bibitem[{{Donati} {et~al.}(1997){Donati}, {Semel}, {Carter}, {Rees}, \&
  {Collier Cameron}}]{1997MNRAS.291..658D}
{Donati}, J.~F., {Semel}, M., {Carter}, B.~D., {Rees}, D.~E., \& {Collier
  Cameron}, A. 1997, \mnras, 291, 658

\bibitem[{{Donati} {et~al.}(1992){Donati}, {Semel}, \&
  {Rees}}]{1992A&A...265..669D}
{Donati}, J.~F., {Semel}, M., \& {Rees}, D.~E. 1992, \aap, 265, 669

\bibitem[{{Dziembowski} \& {Goode}(1996)}]{1996ApJ...458..338D}
{Dziembowski}, W.~A. \& {Goode}, P.~R. 1996, \apj, 458, 338

\bibitem[{{Erspamer} \& {North}(2003)}]{2003A&A...398.1121E}
{Erspamer}, D. \& {North}, P. 2003, \aap, 398, 1121

\bibitem[{{Folsom} {et~al.}(2018){Folsom}, {Bouvier}, {Petit}, {L{\`e}bre},
  {Amard}, {Palacios}, {Morin}, {Donati}, \& {Vidotto}}]{2018MNRAS.474.4956F}
{Folsom}, C.~P., {Bouvier}, J., {Petit}, P., {et~al.} 2018, \mnras, 474, 4956

\bibitem[{{Gaia Collaboration} {et~al.}(2018){Gaia Collaboration}, {Brown},
  {Vallenari}, {Prusti}, {de Bruijne}, {Babusiaux}, {Bailer-Jones}, {Biermann},
  {Evans}, {Eyer}, {Jansen}, {Jordi}, {Klioner}, {Lammers}, {Lindegren},
  {Luri}, {Mignard}, {Panem}, {Pourbaix}, {Randich}, {Sartoretti}, {Siddiqui},
  {Soubiran}, {van Leeuwen}, {Walton}, {Arenou}, {Bastian}, {Cropper},
  {Drimmel}, {Katz}, {Lattanzi}, {Bakker}, {Cacciari}, {Casta{\~n}eda},
  {Chaoul}, {Cheek}, {De Angeli}, {Fabricius}, {Guerra}, {Holl}, {Masana},
  {Messineo}, {Mowlavi}, {Nienartowicz}, {Panuzzo}, {Portell}, {Riello},
  {Seabroke}, {Tanga}, {Th{\'e}venin}, {Gracia-Abril}, {Comoretto},
  {Garcia-Reinaldos}, {Teyssier}, {Altmann}, {Andrae}, {Audard},
  {Bellas-Velidis}, {Benson}, {Berthier}, {Blomme}, {Burgess}, {Busso},
  {Carry}, {Cellino}, {Clementini}, {Clotet}, {Creevey}, {Davidson}, {De
  Ridder}, {Delchambre}, {Dell'Oro}, {Ducourant},
  {Fern{\'a}ndez-Hern{\'a}ndez}, {Fouesneau}, {Fr{\'e}mat}, {Galluccio},
  {Garc{\'\i}a-Torres}, {Gonz{\'a}lez-N{\'u}{\~n}ez}, {Gonz{\'a}lez-Vidal},
  {Gosset}, {Guy}, {Halbwachs}, {Hambly}, {Harrison}, {Hern{\'a}ndez},
  {Hestroffer}, {Hodgkin}, {Hutton}, {Jasniewicz}, {Jean-Antoine-Piccolo},
  {Jordan}, {Korn}, {Krone-Martins}, {Lanzafame}, {Lebzelter}, {L{\"o}ffler},
  {Manteiga}, {Marrese}, {Mart{\'\i}n-Fleitas}, {Moitinho}, {Mora}, {Muinonen},
  {Osinde}, {Pancino}, {Pauwels}, {Petit}, {Recio-Blanco}, {Richards},
  {Rimoldini}, {Robin}, {Sarro}, {Siopis}, {Smith}, {Sozzetti}, {S{\"u}veges},
  {Torra}, {van Reeven}, {Abbas}, {Abreu Aramburu}, {Accart}, {Aerts},
  {Altavilla}, {{\'A}lvarez}, {Alvarez}, {Alves}, {Anderson}, {Andrei},
  {Anglada Varela}, {Antiche}, {Antoja}, {Arcay}, {Astraatmadja}, {Bach},
  {Baker}, {Balaguer-N{\'u}{\~n}ez}, {Balm}, {Barache}, {Barata}, {Barbato},
  {Barblan}, {Barklem}, {Barrado}, {Barros}, {Barstow}, {Bartholom{\'e}
  Mu{\~n}oz}, {Bassilana}, {Becciani}, {Bellazzini}, {Berihuete}, {Bertone},
  {Bianchi}, {Bienaym{\'e}}, {Blanco-Cuaresma}, {Boch}, {Boeche}, {Bombrun},
  {Borrachero}, {Bossini}, {Bouquillon}, {Bourda}, {Bragaglia}, {Bramante},
  {Breddels}, {Bressan}, {Brouillet}, {Br{\"u}semeister}, {Brugaletta},
  {Bucciarelli}, {Burlacu}, {Busonero}, {Butkevich}, {Buzzi}, {Caffau},
  {Cancelliere}, {Cannizzaro}, {Cantat-Gaudin}, {Carballo}, {Carlucci},
  {Carrasco}, {Casamiquela}, {Castellani}, {Castro-Ginard}, {Charlot},
  {Chemin}, {Chiavassa}, {Cocozza}, {Costigan}, {Cowell}, {Crifo}, {Crosta},
  {Crowley}, {Cuypers}, {Dafonte}, {Damerdji}, {Dapergolas}, {David}, {David},
  {de Laverny}, {De Luise}, {De March}, {de Martino}, {de Souza}, {de Torres},
  {Debosscher}, {del Pozo}, {Delbo}, {Delgado}, {Delgado}, {Di Matteo},
  {Diakite}, {Diener}, {Distefano}, {Dolding}, {Drazinos}, {Dur{\'a}n},
  {Edvardsson}, {Enke}, {Eriksson}, {Esquej}, {Eynard Bontemps}, {Fabre},
  {Fabrizio}, {Faigler}, {Falc{\~a}o}, {Farr{\`a}s Casas}, {Federici},
  {Fedorets}, {Fernique}, {Figueras}, {Filippi}, {Findeisen}, {Fonti},
  {Fraile}, {Fraser}, {Fr{\'e}zouls}, {Gai}, {Galleti}, {Garabato},
  {Garc{\'\i}a-Sedano}, {Garofalo}, {Garralda}, {Gavel}, {Gavras}, {Gerssen},
  {Geyer}, {Giacobbe}, {Gilmore}, {Girona}, {Giuffrida}, {Glass}, {Gomes},
  {Granvik}, {Gueguen}, {Guerrier}, {Guiraud}, {Guti{\'e}rrez-S{\'a}nchez},
  {Haigron}, {Hatzidimitriou}, {Hauser}, {Haywood}, {Heiter}, {Helmi}, {Heu},
  {Hilger}, {Hobbs}, {Hofmann}, {Holland}, {Huckle}, {Hypki}, {Icardi},
  {Jan{\ss}en}, {Jevardat de Fombelle}, {Jonker}, {Juh{\'a}sz}, {Julbe},
  {Karampelas}, {Kewley}, {Klar}, {Kochoska}, {Kohley}, {Kolenberg},
  {Kontizas}, {Kontizas}, {Koposov}, {Kordopatis}, {Kostrzewa-Rutkowska},
  {Koubsky}, {Lambert}, {Lanza}, {Lasne}, {Lavigne}, {Le Fustec}, {Le
  Poncin-Lafitte}, {Lebreton}, {Leccia}, {Leclerc}, {Lecoeur-Taibi},
  {Lenhardt}, {Leroux}, {Liao}, {Licata}, {Lindstr{\o}m}, {Lister}, {Livanou},
  {Lobel}, {L{\'o}pez}, {Managau}, {Mann}, {Mantelet}, {Marchal}, {Marchant},
  {Marconi}, {Marinoni}, {Marschalk{\'o}}, {Marshall}, {Martino}, {Marton},
  {Mary}, {Massari}, {Matijevi{\v{c}}}, {Mazeh}, {McMillan}, {Messina},
  {Michalik}, {Millar}, {Molina}, {Molinaro}, {Moln{\'a}r}, {Montegriffo},
  {Mor}, {Morbidelli}, {Morel}, {Morris}, {Mulone}, {Muraveva}, {Musella},
  {Nelemans}, {Nicastro}, {Noval}, {O'Mullane}, {Ord{\'e}novic},
  {Ord{\'o}{\~n}ez-Blanco}, {Osborne}, {Pagani}, {Pagano}, {Pailler},
  {Palacin}, {Palaversa}, {Panahi}, {Pawlak}, {Piersimoni}, {Pineau}, {Plachy},
  {Plum}, {Poggio}, {Poujoulet}, {Pr{\v{s}}a}, {Pulone}, {Racero}, {Ragaini},
  {Rambaux}, {Ramos-Lerate}, {Regibo}, {Reyl{\'e}}, {Riclet}, {Ripepi}, {Riva},
  {Rivard}, {Rixon}, {Roegiers}, {Roelens}, {Romero-G{\'o}mez}, {Rowell},
  {Royer}, {Ruiz-Dern}, {Sadowski}, {Sagrist{\`a} Sell{\'e}s}, {Sahlmann},
  {Salgado}, {Salguero}, {Sanna}, {Santana-Ros}, {Sarasso}, {Savietto},
  {Schultheis}, {Sciacca}, {Segol}, {Segovia}, {S{\'e}gransan}, {Shih},
  {Siltala}, {Silva}, {Smart}, {Smith}, {Solano}, {Solitro}, {Sordo}, {Soria
  Nieto}, {Souchay}, {Spagna}, {Spoto}, {Stampa}, {Steele},
  {Steidelm{\"u}ller}, {Stephenson}, {Stoev}, {Suess}, {Surdej}, {Szabados},
  {Szegedi-Elek}, {Tapiador}, {Taris}, {Tauran}, {Taylor}, {Teixeira},
  {Terrett}, {Teyssandier}, {Thuillot}, {Titarenko}, {Torra Clotet}, {Turon},
  {Ulla}, {Utrilla}, {Uzzi}, {Vaillant}, {Valentini}, {Valette}, {van Elteren},
  {Van Hemelryck}, {van Leeuwen}, {Vaschetto}, {Vecchiato}, {Veljanoski},
  {Viala}, {Vicente}, {Vogt}, {von Essen}, {Voss}, {Votruba}, {Voutsinas},
  {Walmsley}, {Weiler}, {Wertz}, {Wevers}, {Wyrzykowski}, {Yoldas},
  {{\v{Z}}erjal}, {Ziaeepour}, {Zorec}, {Zschocke}, {Zucker}, {Zurbach}, \&
  {Zwitter}}]{2018AA...616A...1G}
{Gaia Collaboration}, {Brown}, A.~G.~A., {Vallenari}, A., {et~al.} 2018, \aap,
  616, A1

\bibitem[{{Gaia Collaboration} {et~al.}(2016){Gaia Collaboration}, {Prusti},
  {de Bruijne}, {Brown}, {Vallenari}, {Babusiaux}, {Bailer-Jones}, {Bastian},
  {Biermann}, {Evans}, {Eyer}, {Jansen}, {Jordi}, {Klioner}, {Lammers},
  {Lindegren}, {Luri}, {Mignard}, {Milligan}, {Panem}, {Poinsignon},
  {Pourbaix}, {Randich}, {Sarri}, {Sartoretti}, {Siddiqui}, {Soubiran},
  {Valette}, {van Leeuwen}, {Walton}, {Aerts}, {Arenou}, {Cropper}, {Drimmel},
  {H{\o}g}, {Katz}, {Lattanzi}, {O'Mullane}, {Grebel}, {Holland}, {Huc},
  {Passot}, {Bramante}, {Cacciari}, {Casta{\~n}eda}, {Chaoul}, {Cheek}, {De
  Angeli}, {Fabricius}, {Guerra}, {Hern{\'a}ndez}, {Jean-Antoine-Piccolo},
  {Masana}, {Messineo}, {Mowlavi}, {Nienartowicz}, {Ord{\'o}{\~n}ez-Blanco},
  {Panuzzo}, {Portell}, {Richards}, {Riello}, {Seabroke}, {Tanga},
  {Th{\'e}venin}, {Torra}, {Els}, {Gracia-Abril}, {Comoretto},
  {Garcia-Reinaldos}, {Lock}, {Mercier}, {Altmann}, {Andrae}, {Astraatmadja},
  {Bellas-Velidis}, {Benson}, {Berthier}, {Blomme}, {Busso}, {Carry},
  {Cellino}, {Clementini}, {Cowell}, {Creevey}, {Cuypers}, {Davidson}, {De
  Ridder}, {de Torres}, {Delchambre}, {Dell'Oro}, {Ducourant}, {Fr{\'e}mat},
  {Garc{\'\i}a-Torres}, {Gosset}, {Halbwachs}, {Hambly}, {Harrison}, {Hauser},
  {Hestroffer}, {Hodgkin}, {Huckle}, {Hutton}, {Jasniewicz}, {Jordan},
  {Kontizas}, {Korn}, {Lanzafame}, {Manteiga}, {Moitinho}, {Muinonen},
  {Osinde}, {Pancino}, {Pauwels}, {Petit}, {Recio-Blanco}, {Robin}, {Sarro},
  {Siopis}, {Smith}, {Smith}, {Sozzetti}, {Thuillot}, {van Reeven}, {Viala},
  {Abbas}, {Abreu Aramburu}, {Accart}, {Aguado}, {Allan}, {Allasia},
  {Altavilla}, {{\'A}lvarez}, {Alves}, {Anderson}, {Andrei}, {Anglada Varela},
  {Antiche}, {Antoja}, {Ant{\'o}n}, {Arcay}, {Atzei}, {Ayache}, {Bach},
  {Baker}, {Balaguer-N{\'u}{\~n}ez}, {Barache}, {Barata}, {Barbier}, {Barblan},
  {Baroni}, {Barrado y Navascu{\'e}s}, {Barros}, {Barstow}, {Becciani},
  {Bellazzini}, {Bellei}, {Bello Garc{\'\i}a}, {Belokurov}, {Bendjoya},
  {Berihuete}, {Bianchi}, {Bienaym{\'e}}, {Billebaud}, {Blagorodnova},
  {Blanco-Cuaresma}, {Boch}, {Bombrun}, {Borrachero}, {Bouquillon}, {Bourda},
  {Bouy}, {Bragaglia}, {Breddels}, {Brouillet}, {Br{\"u}semeister},
  {Bucciarelli}, {Budnik}, {Burgess}, {Burgon}, {Burlacu}, {Busonero}, {Buzzi},
  {Caffau}, {Cambras}, {Campbell}, {Cancelliere}, {Cantat-Gaudin}, {Carlucci},
  {Carrasco}, {Castellani}, {Charlot}, {Charnas}, {Charvet}, {Chassat},
  {Chiavassa}, {Clotet}, {Cocozza}, {Collins}, {Collins}, {Costigan}, {Crifo},
  {Cross}, {Crosta}, {Crowley}, {Dafonte}, {Damerdji}, {Dapergolas}, {David},
  {David}, {De Cat}, {de Felice}, {de Laverny}, {De Luise}, {De March}, {de
  Martino}, {de Souza}, {Debosscher}, {del Pozo}, {Delbo}, {Delgado},
  {Delgado}, {di Marco}, {Di Matteo}, {Diakite}, {Distefano}, {Dolding}, {Dos
  Anjos}, {Drazinos}, {Dur{\'a}n}, {Dzigan}, {Ecale}, {Edvardsson}, {Enke},
  {Erdmann}, {Escolar}, {Espina}, {Evans}, {Eynard Bontemps}, {Fabre},
  {Fabrizio}, {Faigler}, {Falc{\~a}o}, {Farr{\`a}s Casas}, {Faye}, {Federici},
  {Fedorets}, {Fern{\'a}ndez-Hern{\'a}ndez}, {Fernique}, {Fienga}, {Figueras},
  {Filippi}, {Findeisen}, {Fonti}, {Fouesneau}, {Fraile}, {Fraser}, {Fuchs},
  {Furnell}, {Gai}, {Galleti}, {Galluccio}, {Garabato}, {Garc{\'\i}a-Sedano},
  {Gar{\'e}}, {Garofalo}, {Garralda}, {Gavras}, {Gerssen}, {Geyer}, {Gilmore},
  {Girona}, {Giuffrida}, {Gomes}, {Gonz{\'a}lez-Marcos},
  {Gonz{\'a}lez-N{\'u}{\~n}ez}, {Gonz{\'a}lez-Vidal}, {Granvik}, {Guerrier},
  {Guillout}, {Guiraud}, {G{\'u}rpide}, {Guti{\'e}rrez-S{\'a}nchez}, {Guy},
  {Haigron}, {Hatzidimitriou}, {Haywood}, {Heiter}, {Helmi}, {Hobbs},
  {Hofmann}, {Holl}, {Holland}, {Hunt}, {Hypki}, {Icardi}, {Irwin}, {Jevardat
  de Fombelle}, {Jofr{\'e}}, {Jonker}, {Jorissen}, {Julbe}, {Karampelas},
  {Kochoska}, {Kohley}, {Kolenberg}, {Kontizas}, {Koposov}, {Kordopatis},
  {Koubsky}, {Kowalczyk}, {Krone-Martins}, {Kudryashova}, {Kull}, {Bachchan},
  {Lacoste-Seris}, {Lanza}, {Lavigne}, {Le Poncin-Lafitte}, {Lebreton},
  {Lebzelter}, {Leccia}, {Leclerc}, {Lecoeur-Taibi}, {Lemaitre}, {Lenhardt},
  {Leroux}, {Liao}, {Licata}, {Lindstr{\o}m}, {Lister}, {Livanou}, {Lobel},
  {L{\"o}ffler}, {L{\'o}pez}, {Lopez-Lozano}, {Lorenz}, {Loureiro},
  {MacDonald}, {Magalh{\~a}es Fernandes}, {Managau}, {Mann}, {Mantelet},
  {Marchal}, {Marchant}, {Marconi}, {Marie}, {Marinoni}, {Marrese},
  {Marschalk{\'o}}, {Marshall}, {Mart{\'\i}n-Fleitas}, {Martino}, {Mary},
  {Matijevi{\v{c}}}, {Mazeh}, {McMillan}, {Messina}, {Mestre}, {Michalik},
  {Millar}, {Miranda}, {Molina}, {Molinaro}, {Molinaro}, {Moln{\'a}r},
  {Moniez}, {Montegriffo}, {Monteiro}, {Mor}, {Mora}, {Morbidelli}, {Morel},
  {Morgenthaler}, {Morley}, {Morris}, {Mulone}, {Muraveva}, {Musella},
  {Narbonne}, {Nelemans}, {Nicastro}, {Noval}, {Ord{\'e}novic},
  {Ordieres-Mer{\'e}}, {Osborne}, {Pagani}, {Pagano}, {Pailler}, {Palacin},
  {Palaversa}, {Parsons}, {Paulsen}, {Pecoraro}, {Pedrosa}, {Pentik{\"a}inen},
  {Pereira}, {Pichon}, {Piersimoni}, {Pineau}, {Plachy}, {Plum}, {Poujoulet},
  {Pr{\v{s}}a}, {Pulone}, {Ragaini}, {Rago}, {Rambaux}, {Ramos-Lerate},
  {Ranalli}, {Rauw}, {Read}, {Regibo}, {Renk}, {Reyl{\'e}}, {Ribeiro},
  {Rimoldini}, {Ripepi}, {Riva}, {Rixon}, {Roelens}, {Romero-G{\'o}mez},
  {Rowell}, {Royer}, {Rudolph}, {Ruiz-Dern}, {Sadowski}, {Sagrist{\`a}
  Sell{\'e}s}, {Sahlmann}, {Salgado}, {Salguero}, {Sarasso}, {Savietto},
  {Schnorhk}, {Schultheis}, {Sciacca}, {Segol}, {Segovia}, {Segransan},
  {Serpell}, {Shih}, {Smareglia}, {Smart}, {Smith}, {Solano}, {Solitro},
  {Sordo}, {Soria Nieto}, {Souchay}, {Spagna}, {Spoto}, {Stampa}, {Steele},
  {Steidelm{\"u}ller}, {Stephenson}, {Stoev}, {Suess}, {S{\"u}veges}, {Surdej},
  {Szabados}, {Szegedi-Elek}, {Tapiador}, {Taris}, {Tauran}, {Taylor},
  {Teixeira}, {Terrett}, {Tingley}, {Trager}, {Turon}, {Ulla}, {Utrilla},
  {Valentini}, {van Elteren}, {Van Hemelryck}, {van Leeuwen}, {Varadi},
  {Vecchiato}, {Veljanoski}, {Via}, {Vicente}, {Vogt}, {Voss}, {Votruba},
  {Voutsinas}, {Walmsley}, {Weiler}, {Weingrill}, {Werner}, {Wevers},
  {Whitehead}, {Wyrzykowski}, {Yoldas}, {{\v{Z}}erjal}, {Zucker}, {Zurbach},
  {Zwitter}, {Alecu}, {Allen}, {Allende Prieto}, {Amorim},
  {Anglada-Escud{\'e}}, {Arsenijevic}, {Azaz}, {Balm}, {Beck}, {Bernstein},
  {Bigot}, {Bijaoui}, {Blasco}, {Bonfigli}, {Bono}, {Boudreault}, {Bressan},
  {Brown}, {Brunet}, {Bunclark}, {Buonanno}, {Butkevich}, {Carret}, {Carrion},
  {Chemin}, {Ch{\'e}reau}, {Corcione}, {Darmigny}, {de Boer}, {de Teodoro}, {de
  Zeeuw}, {Delle Luche}, {Domingues}, {Dubath}, {Fodor}, {Fr{\'e}zouls},
  {Fries}, {Fustes}, {Fyfe}, {Gallardo}, {Gallegos}, {Gardiol}, {Gebran},
  {Gomboc}, {G{\'o}mez}, {Grux}, {Gueguen}, {Heyrovsky}, {Hoar}, {Iannicola},
  {Isasi Parache}, {Janotto}, {Joliet}, {Jonckheere}, {Keil}, {Kim},
  {Klagyivik}, {Klar}, {Knude}, {Kochukhov}, {Kolka}, {Kos}, {Kutka}, {Lainey},
  {LeBouquin}, {Liu}, {Loreggia}, {Makarov}, {Marseille}, {Martayan},
  {Martinez-Rubi}, {Massart}, {Meynadier}, {Mignot}, {Munari}, {Nguyen},
  {Nordlander}, {Ocvirk}, {O'Flaherty}, {Olias Sanz}, {Ortiz}, {Osorio},
  {Oszkiewicz}, {Ouzounis}, {Palmer}, {Park}, {Pasquato}, {Peltzer}, {Peralta},
  {P{\'e}turaud}, {Pieniluoma}, {Pigozzi}, {Poels}, {Prat}, {Prod'homme},
  {Raison}, {Rebordao}, {Risquez}, {Rocca-Volmerange}, {Rosen}, {Ruiz-Fuertes},
  {Russo}, {Sembay}, {Serraller Vizcaino}, {Short}, {Siebert}, {Silva},
  {Sinachopoulos}, {Slezak}, {Soffel}, {Sosnowska}, {Strai{\v{z}}ys}, {ter
  Linden}, {Terrell}, {Theil}, {Tiede}, {Troisi}, {Tsalmantza}, {Tur},
  {Vaccari}, {Vachier}, {Valles}, {Van Hamme}, {Veltz}, {Virtanen}, {Wallut},
  {Wichmann}, {Wilkinson}, {Ziaeepour}, \& {Zschocke}}]{2016A&A...595A...1G}
{Gaia Collaboration}, {Prusti}, T., {de Bruijne}, J.~H.~J., {et~al.} 2016,
  \aap, 595, A1

\bibitem[{{Gaia Collaboration} {et~al.}(2023){Gaia Collaboration}, {Vallenari},
  {Brown}, {Prusti}, {de Bruijne}, {Arenou}, {Babusiaux}, {Biermann},
  {Creevey}, {Ducourant}, {Evans}, {Eyer}, {Guerra}, {Hutton}, {Jordi},
  {Klioner}, {Lammers}, {Lindegren}, {Luri}, {Mignard}, {Panem}, {Pourbaix},
  {Randich}, {Sartoretti}, {Soubiran}, {Tanga}, {Walton}, {Bailer-Jones},
  {Bastian}, {Drimmel}, {Jansen}, {Katz}, {Lattanzi}, {van Leeuwen}, {Bakker},
  {Cacciari}, {Casta{\~n}eda}, {De Angeli}, {Fabricius}, {Fouesneau},
  {Fr{\'e}mat}, {Galluccio}, {Guerrier}, {Heiter}, {Masana}, {Messineo},
  {Mowlavi}, {Nicolas}, {Nienartowicz}, {Pailler}, {Panuzzo}, {Riclet}, {Roux},
  {Seabroke}, {Sordo}, {Th{\'e}venin}, {Gracia-Abril}, {Portell}, {Teyssier},
  {Altmann}, {Andrae}, {Audard}, {Bellas-Velidis}, {Benson}, {Berthier},
  {Blomme}, {Burgess}, {Busonero}, {Busso}, {C{\'a}novas}, {Carry}, {Cellino},
  {Cheek}, {Clementini}, {Damerdji}, {Davidson}, {de Teodoro}, {Nu{\~n}ez
  Campos}, {Delchambre}, {Dell'Oro}, {Esquej}, {Fern{\'a}ndez-Hern{\'a}ndez},
  {Fraile}, {Garabato}, {Garc{\'\i}a-Lario}, {Gosset}, {Haigron}, {Halbwachs},
  {Hambly}, {Harrison}, {Hern{\'a}ndez}, {Hestroffer}, {Hodgkin}, {Holl},
  {Jan{\ss}en}, {Jevardat de Fombelle}, {Jordan}, {Krone-Martins}, {Lanzafame},
  {L{\"o}ffler}, {Marchal}, {Marrese}, {Moitinho}, {Muinonen}, {Osborne},
  {Pancino}, {Pauwels}, {Recio-Blanco}, {Reyl{\'e}}, {Riello}, {Rimoldini},
  {Roegiers}, {Rybizki}, {Sarro}, {Siopis}, {Smith}, {Sozzetti}, {Utrilla},
  {van Leeuwen}, {Abbas}, {{\'A}brah{\'a}m}, {Abreu Aramburu}, {Aerts},
  {Aguado}, {Ajaj}, {Aldea-Montero}, {Altavilla}, {{\'A}lvarez}, {Alves},
  {Anders}, {Anderson}, {Anglada Varela}, {Antoja}, {Baines}, {Baker},
  {Balaguer-N{\'u}{\~n}ez}, {Balbinot}, {Balog}, {Barache}, {Barbato},
  {Barros}, {Barstow}, {Bartolom{\'e}}, {Bassilana}, {Bauchet}, {Becciani},
  {Bellazzini}, {Berihuete}, {Bernet}, {Bertone}, {Bianchi}, {Binnenfeld},
  {Blanco-Cuaresma}, {Blazere}, {Boch}, {Bombrun}, {Bossini}, {Bouquillon},
  {Bragaglia}, {Bramante}, {Breedt}, {Bressan}, {Brouillet}, {Brugaletta},
  {Bucciarelli}, {Burlacu}, {Butkevich}, {Buzzi}, {Caffau}, {Cancelliere},
  {Cantat-Gaudin}, {Carballo}, {Carlucci}, {Carnerero}, {Carrasco},
  {Casamiquela}, {Castellani}, {Castro-Ginard}, {Chaoul}, {Charlot}, {Chemin},
  {Chiaramida}, {Chiavassa}, {Chornay}, {Comoretto}, {Contursi}, {Cooper},
  {Cornez}, {Cowell}, {Crifo}, {Cropper}, {Crosta}, {Crowley}, {Dafonte},
  {Dapergolas}, {David}, {David}, {de Laverny}, {De Luise}, {De March}, {De
  Ridder}, {de Souza}, {de Torres}, {del Peloso}, {del Pozo}, {Delbo},
  {Delgado}, {Delisle}, {Demouchy}, {Dharmawardena}, {Di Matteo}, {Diakite},
  {Diener}, {Distefano}, {Dolding}, {Edvardsson}, {Enke}, {Fabre}, {Fabrizio},
  {Faigler}, {Fedorets}, {Fernique}, {Fienga}, {Figueras}, {Fournier},
  {Fouron}, {Fragkoudi}, {Gai}, {Garcia-Gutierrez}, {Garcia-Reinaldos},
  {Garc{\'\i}a-Torres}, {Garofalo}, {Gavel}, {Gavras}, {Gerlach}, {Geyer},
  {Giacobbe}, {Gilmore}, {Girona}, {Giuffrida}, {Gomel}, {Gomez},
  {Gonz{\'a}lez-N{\'u}{\~n}ez}, {Gonz{\'a}lez-Santamar{\'\i}a},
  {Gonz{\'a}lez-Vidal}, {Granvik}, {Guillout}, {Guiraud},
  {Guti{\'e}rrez-S{\'a}nchez}, {Guy}, {Hatzidimitriou}, {Hauser}, {Haywood},
  {Helmer}, {Helmi}, {Sarmiento}, {Hidalgo}, {Hilger}, {H{\l}adczuk}, {Hobbs},
  {Holland}, {Huckle}, {Jardine}, {Jasniewicz}, {Jean-Antoine Piccolo},
  {Jim{\'e}nez-Arranz}, {Jorissen}, {Juaristi Campillo}, {Julbe}, {Karbevska},
  {Kervella}, {Khanna}, {Kontizas}, {Kordopatis}, {Korn}, {K{\'o}sp{\'a}l},
  {Kostrzewa-Rutkowska}, {Kruszy{\'n}ska}, {Kun}, {Laizeau}, {Lambert},
  {Lanza}, {Lasne}, {Le Campion}, {Lebreton}, {Lebzelter}, {Leccia}, {Leclerc},
  {Lecoeur-Taibi}, {Liao}, {Licata}, {Lindstr{\o}m}, {Lister}, {Livanou},
  {Lobel}, {Lorca}, {Loup}, {Madrero Pardo}, {Magdaleno Romeo}, {Managau},
  {Mann}, {Manteiga}, {Marchant}, {Marconi}, {Marcos}, {Marcos Santos},
  {Mar{\'\i}n Pina}, {Marinoni}, {Marocco}, {Marshall}, {Martin Polo},
  {Mart{\'\i}n-Fleitas}, {Marton}, {Mary}, {Masip}, {Massari},
  {Mastrobuono-Battisti}, {Mazeh}, {McMillan}, {Messina}, {Michalik}, {Millar},
  {Mints}, {Molina}, {Molinaro}, {Moln{\'a}r}, {Monari}, {Mongui{\'o}},
  {Montegriffo}, {Montero}, {Mor}, {Mora}, {Morbidelli}, {Morel}, {Morris},
  {Muraveva}, {Murphy}, {Musella}, {Nagy}, {Noval}, {Oca{\~n}a}, {Ogden},
  {Ordenovic}, {Osinde}, {Pagani}, {Pagano}, {Palaversa}, {Palicio},
  {Pallas-Quintela}, {Panahi}, {Payne-Wardenaar}, {Pe{\~n}alosa Esteller},
  {Penttil{\"a}}, {Pichon}, {Piersimoni}, {Pineau}, {Plachy}, {Plum}, {Poggio},
  {Pr{\v{s}}a}, {Pulone}, {Racero}, {Ragaini}, {Rainer}, {Raiteri}, {Rambaux},
  {Ramos}, {Ramos-Lerate}, {Re Fiorentin}, {Regibo}, {Richards}, {Rios Diaz},
  {Ripepi}, {Riva}, {Rix}, {Rixon}, {Robichon}, {Robin}, {Robin}, {Roelens},
  {Rogues}, {Rohrbasser}, {Romero-G{\'o}mez}, {Rowell}, {Royer}, {Ruz Mieres},
  {Rybicki}, {Sadowski}, {S{\'a}ez N{\'u}{\~n}ez}, {Sagrist{\`a} Sell{\'e}s},
  {Sahlmann}, {Salguero}, {Samaras}, {Sanchez Gimenez}, {Sanna},
  {Santove{\~n}a}, {Sarasso}, {Schultheis}, {Sciacca}, {Segol}, {Segovia},
  {S{\'e}gransan}, {Semeux}, {Shahaf}, {Siddiqui}, {Siebert}, {Siltala},
  {Silvelo}, {Slezak}, {Slezak}, {Smart}, {Snaith}, {Solano}, {Solitro},
  {Souami}, {Souchay}, {Spagna}, {Spina}, {Spoto}, {Steele},
  {Steidelm{\"u}ller}, {Stephenson}, {S{\"u}veges}, {Surdej}, {Szabados},
  {Szegedi-Elek}, {Taris}, {Taylor}, {Teixeira}, {Tolomei}, {Tonello}, {Torra},
  {Torra}, {Torralba Elipe}, {Trabucchi}, {Tsounis}, {Turon}, {Ulla}, {Unger},
  {Vaillant}, {van Dillen}, {van Reeven}, {Vanel}, {Vecchiato}, {Viala},
  {Vicente}, {Voutsinas}, {Weiler}, {Wevers}, {Wyrzykowski}, {Yoldas}, {Yvard},
  {Zhao}, {Zorec}, {Zucker}, \& {Zwitter}}]{2023A&A...674A...1G}
{Gaia Collaboration}, {Vallenari}, A., {Brown}, A.~G.~A., {et~al.} 2023, \aap,
  674, A1

\bibitem[{{Ghazaryan} {et~al.}(2018){Ghazaryan}, {Alecian}, \&
  {Hakobyan}}]{2018MNRAS.480.2953G}
{Ghazaryan}, S., {Alecian}, G., \& {Hakobyan}, A.~A. 2018, \mnras, 480, 2953

\bibitem[{{Gray} {et~al.}(2016){Gray}, {Corbally}, {De Cat}, {Fu}, {Ren},
  {Shi}, {Luo}, {Zhang}, {Wu}, {Cao}, {Li}, {Zhang}, {Hou}, \&
  {Wang}}]{2016AJ....151...13G}
{Gray}, R.~O., {Corbally}, C.~J., {De Cat}, P., {et~al.} 2016, \aj, 151, 13

\bibitem[{{Grevesse} \& {Sauval}(1998)}]{1998SSRv...85..161G}
{Grevesse}, N. \& {Sauval}, A.~J. 1998, \ssr, 85, 161

\bibitem[{{Holdsworth} \& {Brunsden}(2020)}]{2020PASP..132j5001H}
{Holdsworth}, D.~L. \& {Brunsden}, E. 2020, \pasp, 132, 105001

\bibitem[{{Holdsworth} {et~al.}(2016){Holdsworth}, {Kurtz}, {Smalley}, {Saio},
  {Handler}, {Murphy}, \& {Lehmann}}]{2016MNRAS.462..876H}
{Holdsworth}, D.~L., {Kurtz}, D.~W., {Smalley}, B., {et~al.} 2016, \mnras, 462,
  876

\bibitem[{{Holdsworth} {et~al.}(2018){Holdsworth}, {Saio}, {Bowman}, {Kurtz},
  {Sefako}, {Joyce}, {Lambert}, \& {Smalley}}]{2018MNRAS.476..601H}
{Holdsworth}, D.~L., {Saio}, H., {Bowman}, D.~M., {et~al.} 2018, \mnras, 476,
  601

\bibitem[{{Huber} {et~al.}(2014){Huber}, {Silva Aguirre}, {Matthews},
  {Pinsonneault}, {Gaidos}, {Garc{\'\i}a}, {Hekker}, {Mathur}, {Mosser},
  {Torres}, {Bastien}, {Basu}, {Bedding}, {Chaplin}, {Demory}, {Fleming},
  {Guo}, {Mann}, {Rowe}, {Serenelli}, {Smith}, \&
  {Stello}}]{2014ApJS..211....2H}
{Huber}, D., {Silva Aguirre}, V., {Matthews}, J.~M., {et~al.} 2014, \apjs, 211,
  2

\bibitem[{{H{\"u}mmerich} {et~al.}(2018){H{\"u}mmerich}, {Mikul{\'a}{\v{s}}ek},
  {Paunzen}, {Bernhard}, {Jan{\'\i}k}, {Yakunin}, {Pribulla}, {Va{\v{n}}ko}, \&
  {Mat{\v{e}}chov{\'a}}}]{H_mmerich_2018}
{H{\"u}mmerich}, S., {Mikul{\'a}{\v{s}}ek}, Z., {Paunzen}, E., {et~al.} 2018,
  \aap, 619, A98

\bibitem[{{Koch} {et~al.}(2010){Koch}, {Borucki}, {Basri}, {Batalha}, {Brown},
  {Caldwell}, {Christensen-Dalsgaard}, {Cochran}, {DeVore}, {Dunham},
  {Gautier}, {Geary}, {Gilliland}, {Gould}, {Jenkins}, {Kondo}, {Latham},
  {Lissauer}, {Marcy}, {Monet}, {Sasselov}, {Boss}, {Brownlee}, {Caldwell},
  {Dupree}, {Howell}, {Kjeldsen}, {Meibom}, {Morrison}, {Owen}, {Reitsema},
  {Tarter}, {Bryson}, {Dotson}, {Gazis}, {Haas}, {Kolodziejczak}, {Rowe}, {Van
  Cleve}, {Allen}, {Chandrasekaran}, {Clarke}, {Li}, {Quintana}, {Tenenbaum},
  {Twicken}, \& {Wu}}]{2010ApJ...713L..79K}
{Koch}, D.~G., {Borucki}, W.~J., {Basri}, G., {et~al.} 2010, \apjl, 713, L79

\bibitem[{{Kochukhov}(2003)}]{2003A&A...404..669K}
{Kochukhov}, O. 2003, \aap, 404, 669

\bibitem[{{Luo} {et~al.}(2022){Luo}, {Zhao}, {Zhao}, \& {et
  al.}}]{2022yCat.5156....0L}
{Luo}, A.~L., {Zhao}, Y.~H., {Zhao}, G., \& {et al.} 2022, VizieR Online Data
  Catalog, V/156

\bibitem[{{Mathys}(1989)}]{1989FCPh...13..143M}
{Mathys}, G. 1989, \fcp, 13, 143

\bibitem[{{Matsuno} {et~al.}(2022){Matsuno}, {Starkenburg}, {Balbinot}, \&
  {Helmi}}]{2022arXiv221211639M}
{Matsuno}, T., {Starkenburg}, E., {Balbinot}, E., \& {Helmi}, A. 2022, arXiv
  e-prints, arXiv:2212.11639

\bibitem[{{Pinsonneault} {et~al.}(2012){Pinsonneault}, {An},
  {Molenda-{\.Z}akowicz}, {Chaplin}, {Metcalfe}, \&
  {Bruntt}}]{2012ApJS..199...30P}
{Pinsonneault}, M.~H., {An}, D., {Molenda-{\.Z}akowicz}, J., {et~al.} 2012,
  \apjs, 199, 30

\bibitem[{{Preston}(1974)}]{1974ARA&A..12..257P}
{Preston}, G.~W. 1974, \araa, 12, 257

\bibitem[{{Rees} \& {Semel}(1979)}]{1979A&A....74....1R}
{Rees}, D.~E. \& {Semel}, M.~D. 1979, \aap, 74, 1

\bibitem[{{Ricker} {et~al.}(2015){Ricker}, {Winn}, {Vanderspek}, {Latham},
  {Bakos}, {Bean}, {Berta-Thompson}, {Brown}, {Buchhave}, {Butler}, {Butler},
  {Chaplin}, {Charbonneau}, {Christensen-Dalsgaard}, {Clampin}, {Deming},
  {Doty}, {De Lee}, {Dressing}, {Dunham}, {Endl}, {Fressin}, {Ge}, {Henning},
  {Holman}, {Howard}, {Ida}, {Jenkins}, {Jernigan}, {Johnson}, {Kaltenegger},
  {Kawai}, {Kjeldsen}, {Laughlin}, {Levine}, {Lin}, {Lissauer}, {MacQueen},
  {Marcy}, {McCullough}, {Morton}, {Narita}, {Paegert}, {Palle}, {Pepe},
  {Pepper}, {Quirrenbach}, {Rinehart}, {Sasselov}, {Sato}, {Seager},
  {Sozzetti}, {Stassun}, {Sullivan}, {Szentgyorgyi}, {Torres}, {Udry}, \&
  {Villasenor}}]{2015JATIS...1a4003R}
{Ricker}, G.~R., {Winn}, J.~N., {Vanderspek}, R., {et~al.} 2015, Journal of
  Astronomical Telescopes, Instruments, and Systems, 1, 014003

\bibitem[{Ryabchikova {et~al.}(2015)Ryabchikova, Piskunov, Kurucz, Stempels,
  Heiter, Pakhomov, \& Barklem}]{Ryabchikova_2015}
Ryabchikova, T., Piskunov, N., Kurucz, R.~L., {et~al.} 2015, Physica Scripta,
  90, 054005

\bibitem[{{Saio}(2005)}]{2005MNRAS.360.1022S}
{Saio}, H. 2005, \mnras, 360, 1022

\bibitem[{{Saio} \& {Gautschy}(2004)}]{2004MNRAS.350..485S}
{Saio}, H. \& {Gautschy}, A. 2004, \mnras, 350, 485

\bibitem[{{Shi} {et~al.}(2020){Shi}, {Kurtz}, {Saio}, {Fu}, \&
  {Zhang}}]{2020ApJ...901...15S}
{Shi}, F., {Kurtz}, D., {Saio}, H., {Fu}, J., \& {Zhang}, H. 2020, \apj, 901,
  15

\bibitem[{Sneden {et~al.}(2012)Sneden, Bean, Ivans, Lucatello, \&
  Sobeck}]{Sneden2012}
Sneden, C., Bean, J., Ivans, I., Lucatello, S., \& Sobeck, J. 2012,
  Astrophysics Source Code Library, 02009

\bibitem[{{Stibbs}(1950)}]{1950MNRAS.110..395S}
{Stibbs}, D.~W.~N. 1950, \mnras, 110, 395

\bibitem[{{Xiang} {et~al.}(2022){Xiang}, {Rix}, {Ting}, {Kudritzki}, {Conroy},
  {Zari}, {Shi}, {Przybilla}, {Ramirez-Tannus}, {Tkachenko}, {Gebruers}, \&
  {Liu}}]{2022AA...662A..66X}
{Xiang}, M., {Rix}, H.-W., {Ting}, Y.-S., {et~al.} 2022, \aap, 662, A66

\bibitem[{{Zhao} {et~al.}(2012){Zhao}, {Zhao}, {Chu}, {Jing}, \&
  {Deng}}]{2012RAA....12..723Z}
{Zhao}, G., {Zhao}, Y.-H., {Chu}, Y.-Q., {Jing}, Y.-P., \& {Deng}, L.-C. 2012,
  Research in Astronomy and Astrophysics, 12, 723

\end{thebibliography}
\end{document}